\documentclass[numbers=noenddot]{jfm}
\usepackage[latin9]{inputenc}

\usepackage{array, ragged2e}
\usepackage{float}
\usepackage{mathtools}
\usepackage{graphicx}
\usepackage{esint}
\usepackage{verbatim}
\usepackage{caption}
\usepackage{subcaption}
\renewcommand\[{\begin{equation}}
\renewcommand\]{\end{equation}}
\newcommand{\un}[1]{\,\mathrm{#1}}
\numberwithin{equation}{section}
\usepackage{amsmath} 
\newcommand{\beq}{\begin{equation}}
\newcommand{\eeq}{\end{equation}}
\newcommand{\lb}{\left(}
\newcommand{\rb}{\right)}
\usepackage{tikz}
\usepackage{pgfplots}
\usepackage{overpic}
\usepackage{todonotes}
\author{G.P. Benham \aff{1}\corresp{\email{graham.benham@ladhyx.polytechnique.fr}}, 
J.P. Boucher \aff{1},  
R. Labbé \aff{1},
M. Benzaquen \aff{1},
C. Clanet \aff{1}}
\affiliation{\aff{1} LadHyX, UMR CNRS 7646, Ecole polytechnique, 91128 Palaiseau, France}

\begin{document}

\title{Wave drag on asymmetric bodies}

\maketitle

\abstract{
An asymmetric body with a sharp leading edge and a rounded trailing edge produces a smaller wave disturbance moving forwards than backwards, and this is reflected in the wave drag coefficient.
This  experimental fact is not captured by Michell's theory for wave drag \citep{michell1898xi}.
In this study, we use a tow-tank experiment to investigate the effects of asymmetry on wave drag, and show that these effects can be replicated by modifying Michell's theory to include the growth of a symmetry-breaking boundary layer.
We show that asymmetry can have either a positive or a negative effect on drag, depending on the depth of motion and the Froude number.}

\section{Introduction}

Many existing studies use the inviscid theories of \citet{michell1898xi} and \citet{havelock1919wave,havelock1932theory} to investigate the optimum design of ship hulls \citep{zakerdoost2013ship,zhao2015ship,dambrine2016theoretical,boucher2018thin}. However, asymmetric hull shapes are not addressed, since there is no well-accepted predictive theory for the effect of asymmetry on wave resistance. 
Some studies have shown that viscosity plays an important role in the wave resistance of ship hulls \citep{gotman2002study,lazauskas2009resistance}. In particular, it is argued that the development of the turbulent boundary layer and its detachment point, where applicable, is crucial. 
It is well known that the development of a boundary layer on an asymmetric body is different depending on the direction of motion, due to the dependence of the boundary layer growth rate on the streamwise pressure gradients.
This indicates that a viscous description of the flow is a possible way of studying asymmetric effects, although this is not addressed explicitly in any of the above studies.

More than a century ago, Michell derived the integral formula for the wave resistance on a body, using the approximation of a slender body in an irrotational, inviscid fluid \citep{michell1898xi}. The major shortcoming of this formula is that, due to the reversibility of the steady potential flow formulation, it does not distinguish the difference in wave drag when an asymmetric object moves forwards or backwards. 
However, it is clear that a large number of boats are designed with an asymmetric shape that is more pointed at the front than at the rear, precisely to reduce the wave disturbance. Hence, this theory cannot be used to reliably test design spaces.

Another commonly used method for estimating the wave drag on a body is the formula derived by \citet{havelock1919wave, havelock1932theory}. This approach, which also makes the assumption of an irrotational and inviscid fluid, requires knowledge of the pressure disturbance along the walls of the body. Hence, for an asymmetric body, if the difference in the pressure disturbance between forward and backward motion is known, then this formulation can capture the effects of asymmetry on wave drag. However, in practice it is very difficult to have \textit{a priori} knowledge of the pressure distribution for a given body shape \citep{boucher2018thesis}, which is why, despite the failure to capture asymmetric effects, the Michell formula is much the more popular.

In the present study, we first show the effects of asymmetry experimentally and then discuss how such effects can be predicted theoretically using either computational fluid dynamics, or our new proposed modification to Michell's theory which includes the development of a turbulent boundary layer.

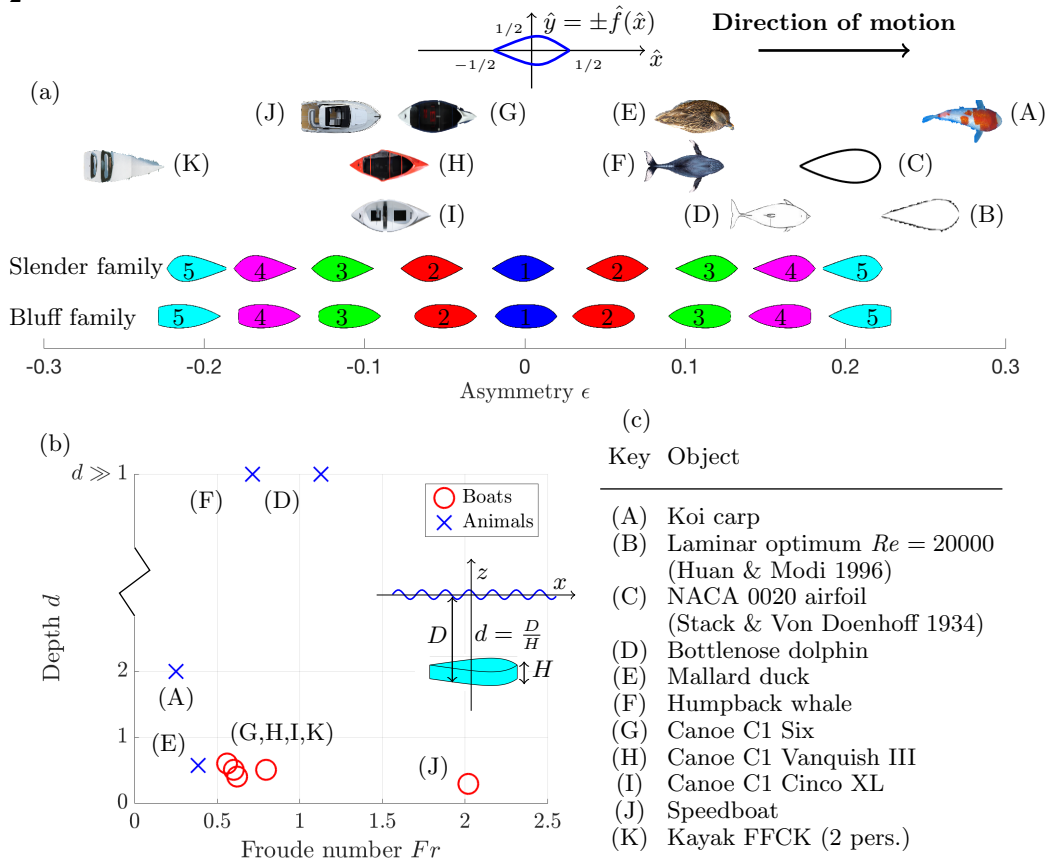
\begin{figure}
\centering
\vspace{1.2cm}
\begin{overpic}[width=1\textwidth]{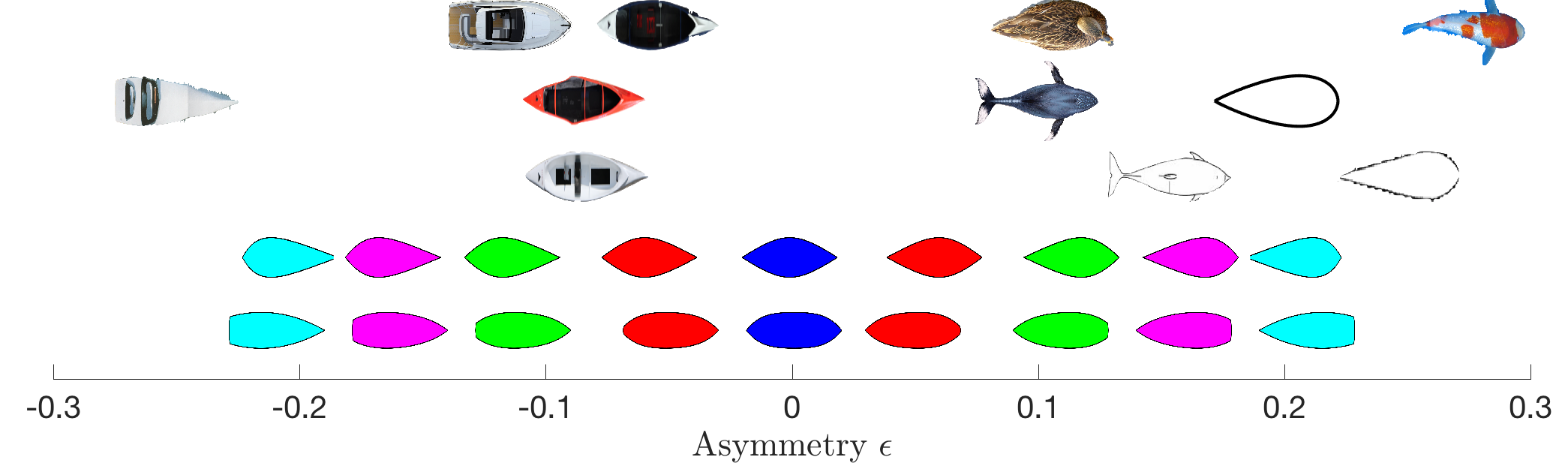}
\put (40, 31) {\begin{tikzpicture}[scale=0.5]
\draw[line width=0.5,->] (-3,0) -- (3,0);
\draw[line width=0.5,->] (0,-1) -- (0,1);
\draw[line width=1,blue] (-1,0) .. controls (0.2,0.5) ..  (1,0);
\draw[line width=1,blue] (-1,0) .. controls (0.2,-0.5) ..  (1,0);
\node at (1.8,0.8){$\hat{y}=\pm \hat{f}(\hat{x})$};
\node at (-0.6,0.55){\tiny $1/2$};
\node at (3.3,-0.2) {$\hat{x}$};
\node at (1.5,-0.4) {\tiny  $1/2$};
\node at (-1.5,-0.4) {\tiny $-1/2$};
\draw[line width=1,->] (6,0) -- (10,0);
\node at (8,0.75){\bf Direction of motion};
\end{tikzpicture}}
\put (98,28) {(A)}
\put (94,18) {(B)}
\put (87,23) {(C)}
\put (59,28) {(E)}
\put (58,23) {(F)}
\put (66,18) {(D)}
\put (47,28) {(G)}
\put (42,23) {(H)}
\put (42,18) {(I)}
\put (24,28) {(J)}
\put (16,23) {(K)}
\put (0,8) {Bluff family}
\put (16,8) {5}
\put (24,8) {4}
\put (32,8) {3}
\put (42,8) {2}
\put (50,8) {1}
\put (58,8) {2}
\put (67,8) {3}
\put (75,8) {4}
\put (84,8) {5}
\put (0,13) {Slender family}
\put (17,12.5) {5}
\put (24,12.5) {4}
\put (32,12.5) {3}
\put (41,12.5) {2}
\put (50,12.5) {1}
\put (59,12.5) {2}
\put (68,12.5) {3}
\put (76,12.5) {4}
\put (83,12.5) {5}
\put (2,30) {(a)}
\put (60,-2) {(c)}
\end{overpic}
\\
\vspace{0.2cm}
\begin{subfigure}{0.55\textwidth}
\begin{tikzpicture}[scale=0.5]
\node at (0,0){\includegraphics[width=1\textwidth]{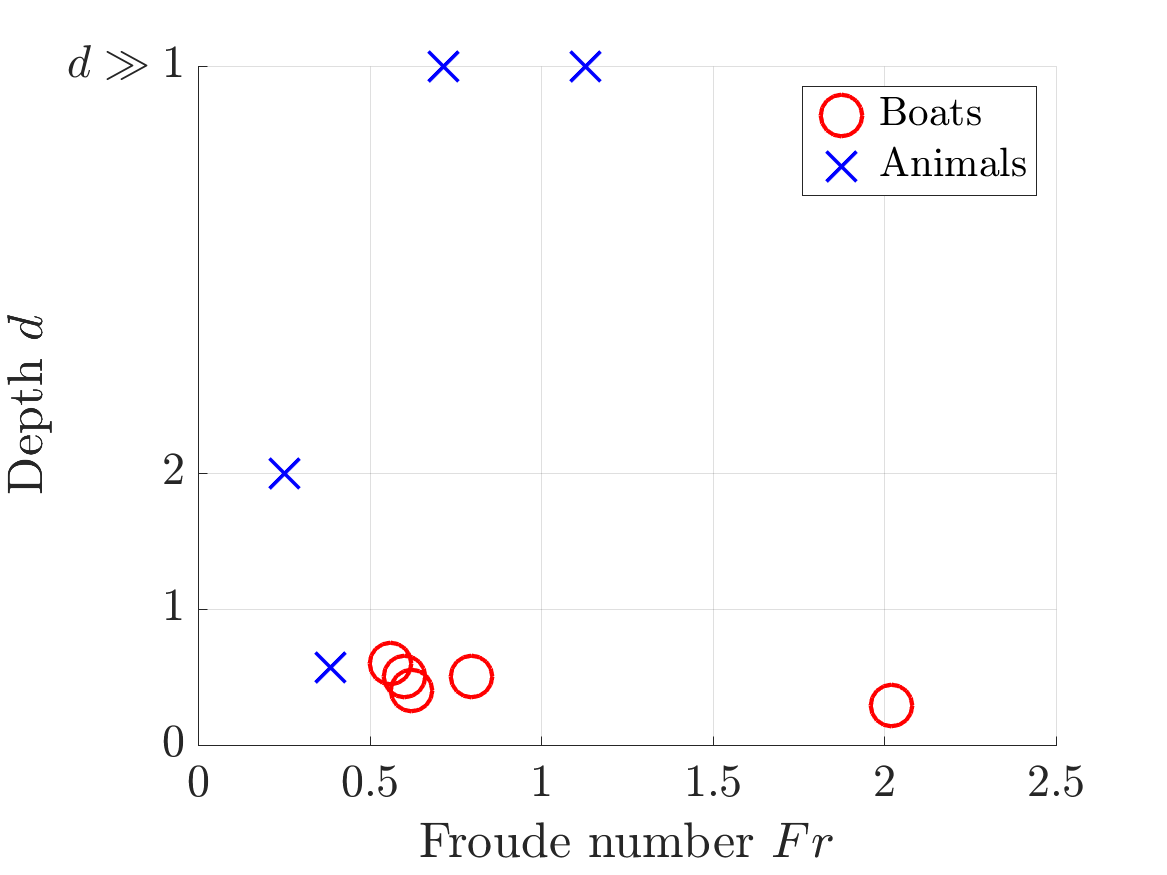}};
\begin{axis}[domain=0:50,hide axis, scale only axis, width=0.7\textwidth,
     height=0.04\textwidth, samples=100, at={(0.2\textwidth,0.19\textwidth)}]
      \addplot[mark=none,color=blue,very thick]{sin(50*x)};
    \end{axis}
\node at (4,-0.5){\includegraphics[width=0.2\textwidth]{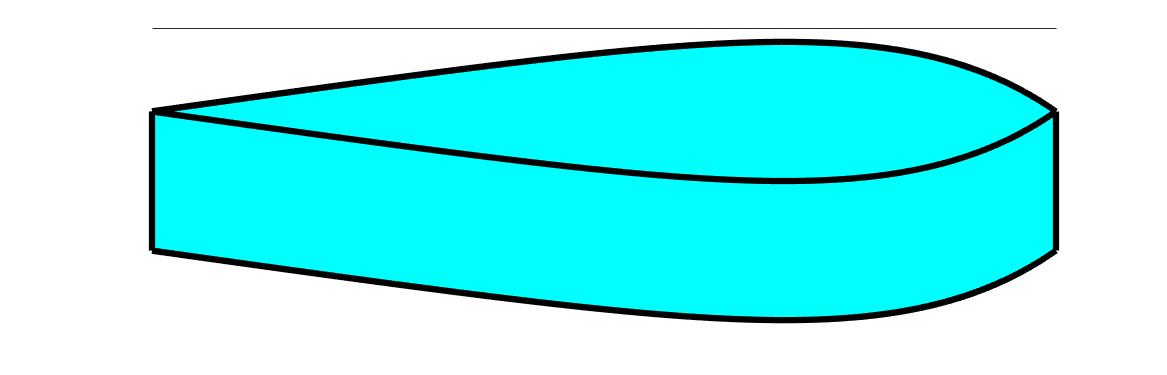}};
%\draw[line width=0.5,->] (4.5,-0.5) -- (6,-0.5);
\draw[line width=0.5,<->] (3.5,-0.75) -- (3.5,1.5);
\draw[line width=0.5,<->] (5.4,-0.8) -- (5.4,-0.2);
\draw[line width=0.3,->] (1.5,1.55) -- (6.75,1.55);
\draw[line width=0.3,->] (4.0,-1.5) -- (4.0,2.5);
\node at (4.3,2.1) {$z$};
\node at (6.35,1.85) {$x$};
\node at (5.9,-0.4){$H$};
\node at (3.1,0.5){$D$};
\node at (5,0.5){$d=\frac{D}{H}$};
\node at (-7,5.5) {(b)};
\node at (-3,4.0) {(F)};
\node at (-1,4.0) {(D)};
\node at (-3.8,-1.2) {(A)};
\node at (-4,-2.4) {(E)};
\node at (-1,-2.1) {(G,H,I,K)};
\node at (3,-3) {(J)};
\draw[white,fill=white] (-5.0,1) rectangle (-4.8,2.8);
\draw[line width=0.5] (-4.9,1) -- (-5.3,1.6) -- (-4.5,2.2) -- (-4.9,2.8);

\end{tikzpicture}
\end{subfigure}
\begin{subfigure}{0.4\textwidth}
\begin{tabular}{cl>{\raggedright}p{1cm}}
%\raggedleft
Key  & Object\\
\hline
(A) &Koi carp\\
(B) &Laminar optimum $\Rey =20000$\\
& \citep{huan1996design}\\
(C) &NACA 0020 airfoil\\
& \citep{stack1934tests}\\
(D) &Bottlenose dolphin\\
(E) &Mallard duck\\
(F) &Humpback whale\\
(G) &Canoe C1 Six \\
(H) &Canoe C1 Vanquish III \\
(I) &Canoe C1 Cinco XL \\
(J) & Speedboat\\
(K) & Kayak FFCK (2 pers.)\\
\end{tabular}
\end{subfigure}
\caption{(a) Examples of different artificial and natural bodies that move near the water surface. All bodies have been scaled to have the same aspect ratio. (b) Non-dimensional average depth of motion $d=D/H$ and Froude number $\mathrm{\it Fr}=U/\sqrt{g L}$ for each of these bodies. (c) Table of the different bodies. \label{existing}}
\end{figure}

\section{The asymmetry parameter $\epsilon$}

\begin{comment}
The difference in wave resistance between forward and backward motion for an asymmetric body can be understood intuitively by considering the leading and trailing edges. Bodies with a leading edge that is long, slender and pointed create a smaller wave disturbance than those with a bluff, or rounded leading edge, due to their ability to \textit{slice} through the water surface. Hence, for a given volume, the body shape that minimises wave drag is asymmetric, with a leading edge sharper than the trailing edge. On the other hand, if the trailing edge is not slender enough, then the flow around the object may be subject to separation, and consequently a larger form drag. In this way, bodies which move near the water surface must have both a pointed leading edge, to reduce wave drag, and a pointed trailing edge, to reduce form drag. But the importance of these two physical phenomena are not necessarily in equal proportion. Indeed, the deeper the motion of the body, the less important its wave drag, and hence its shape should have more of a pointed trailing edge than a leading edge (e.g. a whale). Likewise, bodies with a very shallow motion should have more of a pointed leading edge than a trailing edge (e.g. a dinghy).
\end{comment}

To quantify the asymmetry of a body shape, it is useful to introduce an asymmetry parameter. Throughout this study, for the sake of simplicity, we restrict our attention to shapes which do not vary in the vertical direction. Furthermore, we only consider front-back asymmetry and not asymmetry in the transverse direction.  That is to say, if we take a Cartesian coordinate system with the $x$ direction aligned with the positive direction of motion, and the origin centred at where the object mid-length meets the resting water surface level, then the shape is given by
a function $y=\pm f(x)$ for $ -L/2\leq x\leq L/2,\,-D\leq z \leq -D+H$ and $y=0$ otherwise, where $L$ and $H$ are the body length and height. 
To define a non-dimensional symmetry parameter, it is useful to create non-dimensional variables $\hat{x}=x/L$ and $\hat{f}=f/(2\max \{f(x)\})$ (see figure \ref{existing}(a)). In terms of these new dimensionless variables, we define the asymmetry parameter as the $L_2$ norm 
of the odd function 
\beq
\epsilon =  \kappa \lb{\int_{-1/2}^{1/2} \lb {\hat{f}(\hat{x})-\hat{f}(-\hat{x})} \rb^2 \, \mathrm{d} \hat{x}}\rb^{1/2},
\eeq
where $\kappa=\mathrm{sgn}(\int_{-1/2}^{1/2}\hat{x}\hat{f}(\hat{x})\,\mathrm{d}\hat{x})$ distinguishes the difference between forward and backward motion.

In figure \ref{existing}(a) we compare the value of $\epsilon$ for a variety of natural and artificial bodies that move near the air-water interface. In each case we approximate $\hat{f}(\hat{x})$ as the outline of the plan view of the body (ignoring fins in the case of aquatic creatures), and we ignore variations of the shape with depth. For each object we also compare values of the Froude number $\mathrm{\it Fr}=U/\sqrt{g L}$ and the non-dimensional depth $d=D/H$, where $U$ is a typical velocity scale, $g$ is the gravitational constant, and $D$ is a typical distance between the air-water interface and the deepest part of the body (as illustrated in figure \ref{existing}(b)).

For aquatic creatures with $d\gg 1$, such as the humpback whale or the bottlenose dolphin, we observe positive values of $\epsilon$. This is because, at large depths wave drag is less important than form drag and, hence, for a given body volume, drag is minimised with a streamlined shape with a trailing edge more pointed than its leading edge \citep{videler2012fish}. By contrast, for canoes and other boats with $d\approx 0.5$, we observe negative values of $\epsilon$. This is because at smaller depths, wave drag is more important than form drag and, hence, a pointed leading edge is more preferable. 

Except for the case of the speedboat ($\mathrm{\it Fr}=2.0$), all of the other bodies in figure \ref{existing} have Froude numbers in the range $0.2\leq \mathrm{\it Fr} \leq 1.2$. It is well known that this is the regime where wave drag is typically most significant \citep{michell1898xi,havelock1932theory,tuck1989wave}. Hence, in this study we restrict our attention to Froude numbers in this range and investigate the effect of body asymmetry on drag. Furthermore, we do not investigate the effects of planing \citep{rabaud2014narrow, darmon2014kelvin}.
%, since these only exist at larger Froude numbers \citep{rabaud2014narrow, darmon2014kelvin}.

To investigate the effect of asymmetry, we introduce two families of shapes which have $\epsilon$ values in same range as the existing body shapes, and we number the shapes from 1 to 5, as displayed in figure \ref{existing}. 
One set of shapes is slender, whilst the other is more bluff. The slender family, which we use for the majority of this study, is useful for comparison with Michell's theory, whilst the bluff body is useful for exhibiting the effect of separation and form drag at its largest.
The shapes from both families have the same vertical and horizontal aspect ratios $L/H=3.6$ and $L/W=6$, where $W$ is the body width. The non-dimensional volume $\hat{V}=1/(HLW)\iint f \mathrm{d}x\mathrm{d}z$ is given by $\hat{V}=0.31$ for the slender family and $\hat{V}=0.38$ for the bluff family.
Each family of shapes is defined by an analytical function which is given in Appendix \ref{appA}.

In the next section we use these families of shapes to experimentally investigate the effect of asymmetry on drag at varying depths and Froude numbers. In the subsequent section we replicate the experimental results using a $k$-$\omega$ SST turbulence model, as well as a simple modification of Michell's theory where we account for the development of the turbulent boundary layer. Then we close with a discussion on the effect of the depth of motion, and a summary of all the results.

%\clearpage

\section{Experimental investigation}
\label{sec_exp}

\begin{figure}
\centering
\begin{tikzpicture}[scale=0.65]
\node at (1.9,-0.2) {\includegraphics[width=0.13\textwidth]{figs/3d_hulls_hull_5_new}};
\draw[line width=1] (-3,2) -- (7,2);
\draw[line width=1] (-3,1.8) -- (7,1.8);
\draw[line width=3] (2,1.8) -- (2,1);
\draw[line width=1] (1.25,1) -- (2.75,1);
\draw[line width=1] (1.4,1) -- (1.4,0);
\draw[line width=1] (2.6,1) -- (2.6,0);
\draw[line width=1,->] (2.8,0.8) -- (4,0.8);
%\draw[line width=1,fill=red] (1,-0.5) rectangle (3,0);
\begin{axis}[domain=0:100,hide axis, scale only axis, width=0.55\textwidth,
     height=0.02\textwidth, samples=100, at={(-0.28\textwidth,-0.03\textwidth)}]
      \addplot[mark=none,color=blue,very thick]{sin(50*x)};
    \end{axis}
\draw[line width=0.75,blue,-] (3.0,-0.3) -- (7,-0.3);
\draw[line width=1] (-3.2,1) -- (-3.2,-3) -- (7,-3) -- (7,1);
\draw[line width=1,<->] (-3,-2.3) -- (6.9,-2.3);
\draw[line width=1,<->] (-3.5,-2.9) -- (-3.5,-0.3);
\draw[line width=1,<->] (1,-0.8) -- (3,-0.8);
\draw[line width=1,<->] (3.2,-0.5) -- (3.2,0);
%\draw[line width=1,<->] (0.8,-0.75) -- (0.8,-0.25);
\node at (2,0.42) { Hull};
\node at (-0.5,2.4) { Rail};
\node at (1.1,1.4) { Sensor};
\node at (2.5,-2.6) { $5\un{m}$};
\node at (4.4,0.8) { $U$};
\node at (2.2,-1.2) { $L=0.18\un{m}$};
\node at (4.7,0.1) { $H=0.05\un{m}$};
\node at (8,1) { $ $};
\node at (-4,-1.5) { \rotatebox{90}{$0.25\un{m}$}};
\node at (-3.8,1.75) {(a)};
\end{tikzpicture}
\begin{overpic}[width=0.35\textwidth]{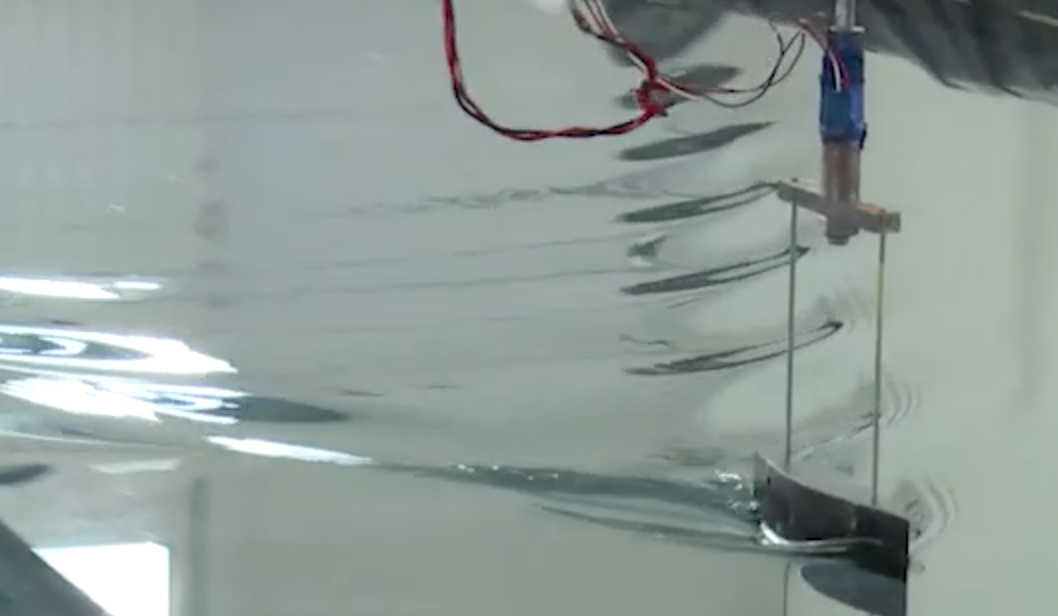}
\put (-10,65) {(b)}
\end{overpic}
\caption{(a) Schematic diagram showing the body hull being pulled through the water at constant velocity by a linear motor and connected via supporting bars to a force sensor. (b) Photograph showing the towed body and its wake pattern.\label{schem}}
\end{figure}

In figure \ref{schem} we display a schematic diagram and a photograph of our experimental set-up. We use a 3D printer to manufacture the two families of asymmetric bodies. These are pulled through the water in a large basin ($5\un{m}\times2\un{m}\times0.25\un{m}$) by a linear motor at constant velocity (in the range $0.4-1.4\un{m/s}$). The hull is connected to the motor by two supporting bars and a force/displacement sensor. Different body depths are achieved using a vertical winch. Experiments are repeated at least $3$ times for accuracy. The dimensions of the hull and water basin are given in the diagram (except for the width of the hull, which is $W=0.03\un{m}$). 

Using this set-up we measure the drag force on the two families of shapes at non-dimensional depths $d=D/H$ between $0.5$ and $2.0$, and at Froude numbers $\mathrm{\it Fr}=U/\sqrt{g L}$ between $0.3$ and $1.0$. We convert the measured force $R$ into a non-dimensional drag coefficient $C_d$ via the relationship
\beq
C_d=\frac{R}{\rho U^2 \Omega^{2/3}},\label{Cdeq}
\eeq
where $\rho$ is the water density and $\Omega=L W  H_w$ is a typical volume scale, where $H_w$ is the wetted depth, which is equal to $H$ when the hull is fully immersed and $d H$ when partly immersed.

\begin{figure}
\centering
\begin{tikzpicture}[scale=0.5]
\node at (0,0) {\begin{overpic}[width=0.45\textwidth]{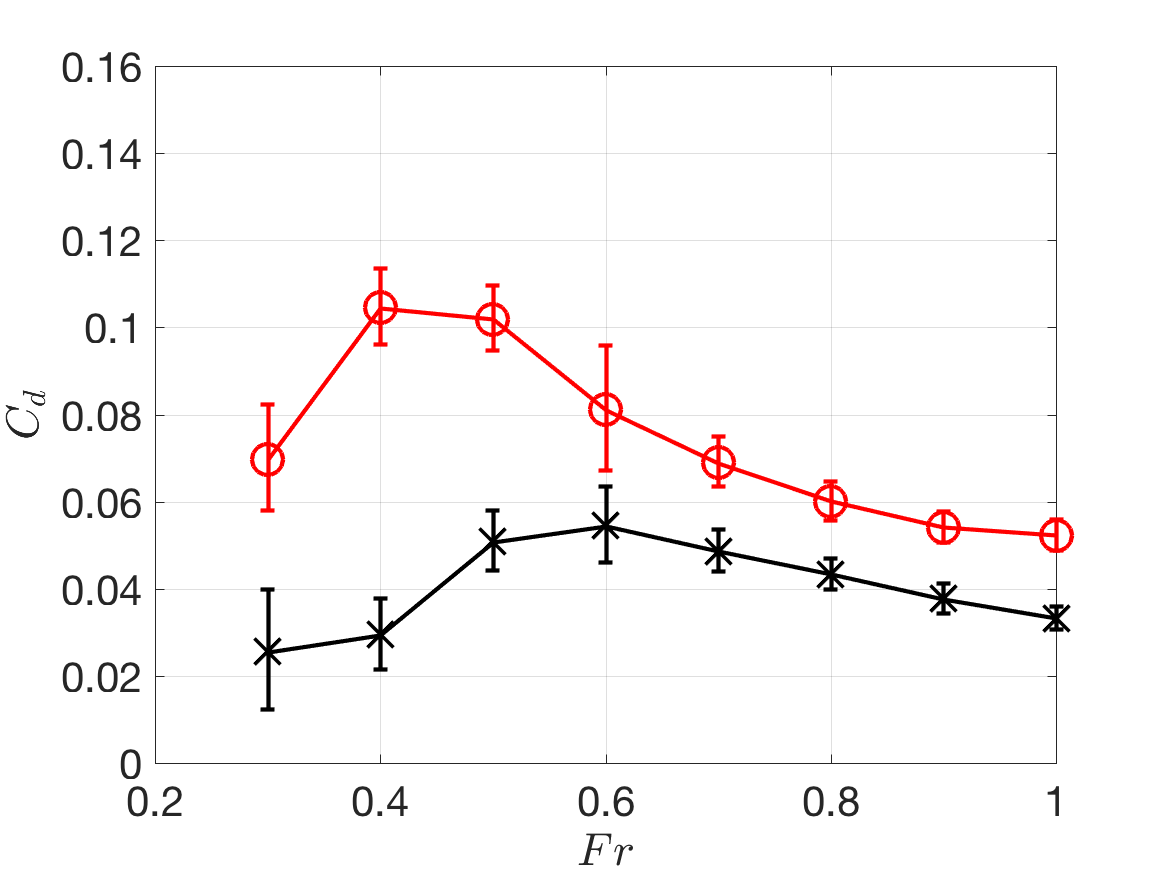}
%\put (35,102) {\bf Depth $\boldsymbol{d=0.5}$}
\put (55,45) {\includegraphics[width=0.15\textwidth]{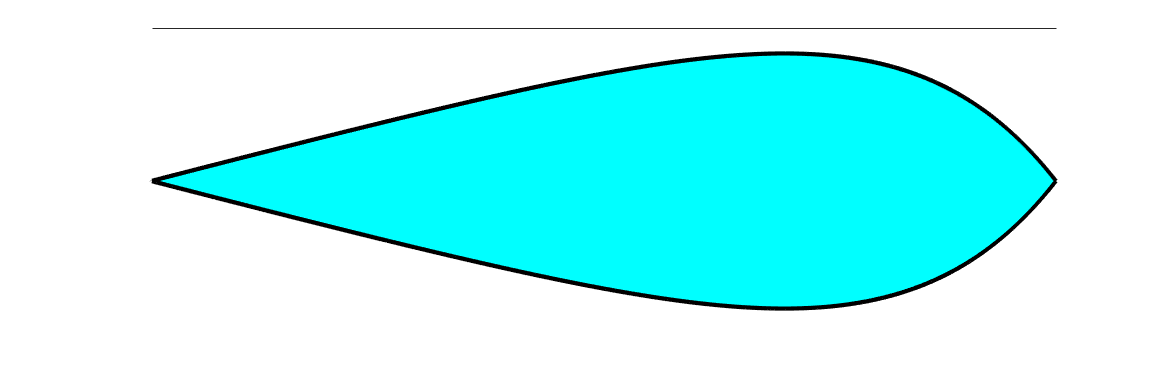}}
\put(70,50){\color{black}\vector(1,0){10}}
\put (0,92) {(a)}
\put (40,12) {\includegraphics[width=0.15\textwidth]{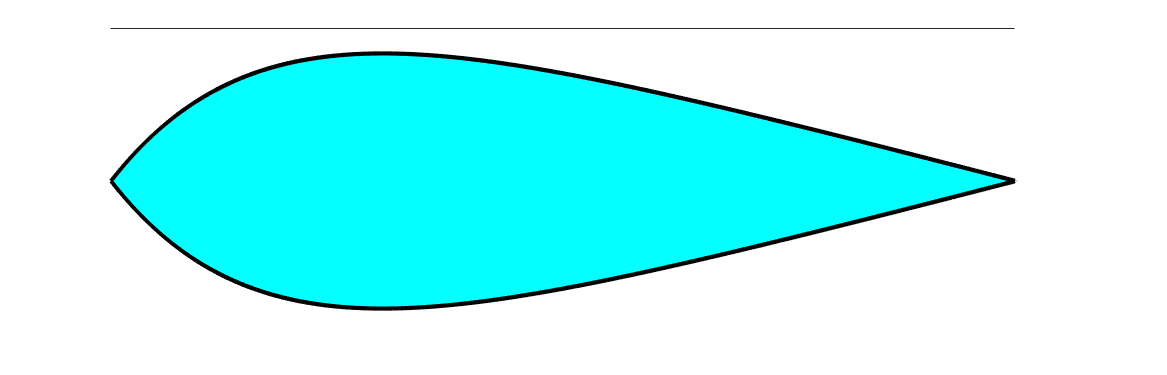}}
\put(50,17){\color{black}\vector(1,0){10}}
\put (5,72) {\includegraphics[width=0.4\textwidth]{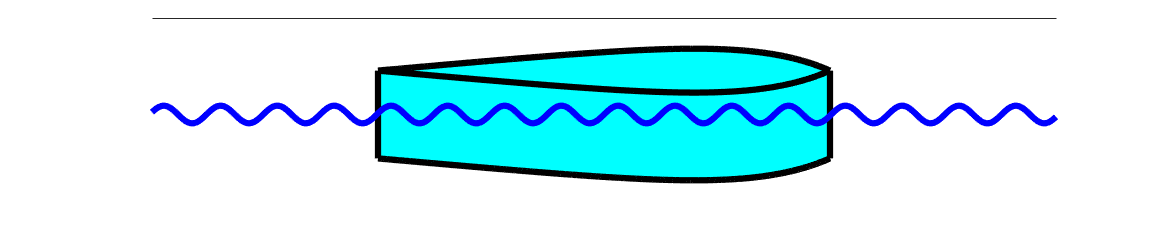}}
\end{overpic}};
\draw[line width=1,->] (2.6,4.4) -- (2.6,4.9);
\draw[line width=1,->] (2.6,6.) -- (2.6,5.5);
\draw[line width=0.5] (2.3,4.9) -- (3.0,4.9);
\draw[line width=1,->] (-2.6,4.4) -- (-2.6,4.9);
\draw[line width=1,->] (-2.6,6.3) -- (-2.6,5.8);
\draw[line width=0.5] (-2.1,4.9) -- (-3.0,4.9);
\draw[line width=0.5] (-2.1,5.8) -- (-3.0,5.8);
\node at (-3.5,4.8){$H$};
\node at (3.4,4.8){$D$};
\node at (0,7.2) {\bf Depth $\boldsymbol{d=D/H=0.5}$};
\node at (-4,3.3) {\includegraphics[width=0.016\textwidth]{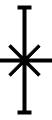}};
\node at (-2.8,3.3){$\epsilon<0$};
\node at (-1.5,3.3) {\includegraphics[width=0.0175\textwidth]{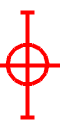}};
\node at (-0.3,3.3){$\epsilon>0$};
\draw[line width=0.25] (-4.3,2.7) rectangle (0.5,3.8);
\end{tikzpicture}
\begin{tikzpicture}[scale=0.5]
\node at (0,0) {\begin{overpic}[width=0.45\textwidth]{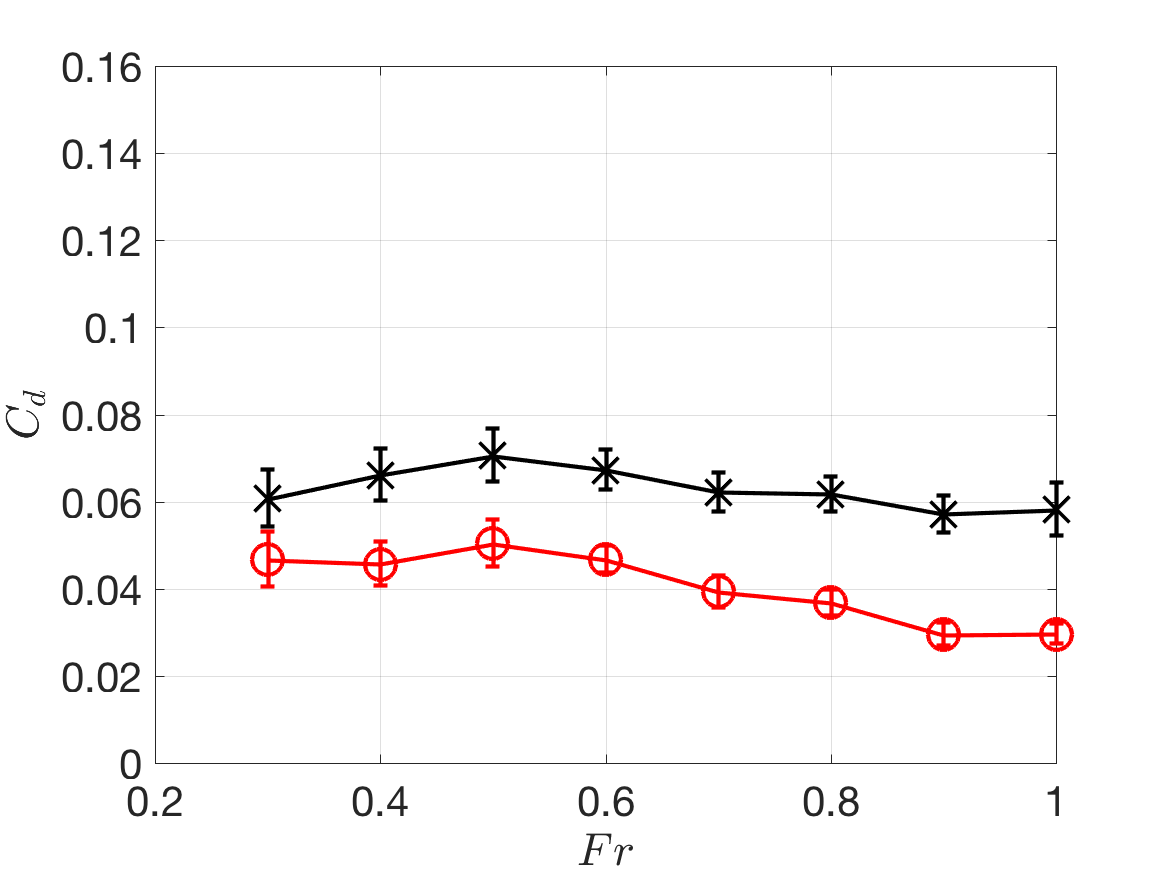}
%\put (35,72) {\bf Depth $\boldsymbol{d=2.0}$}
\put (55,40) {\includegraphics[width=0.15\textwidth]{figs/old_hull_5f}}
\put(65,45){\color{black}\vector(1,0){10}}
\put (40,12) {\includegraphics[width=0.15\textwidth]{figs/old_hull_5b}}
\put(55,17){\color{black}\vector(1,0){10}}
\put (0,92) {(b)}
\put (5,72) {\includegraphics[width=0.4\textwidth]{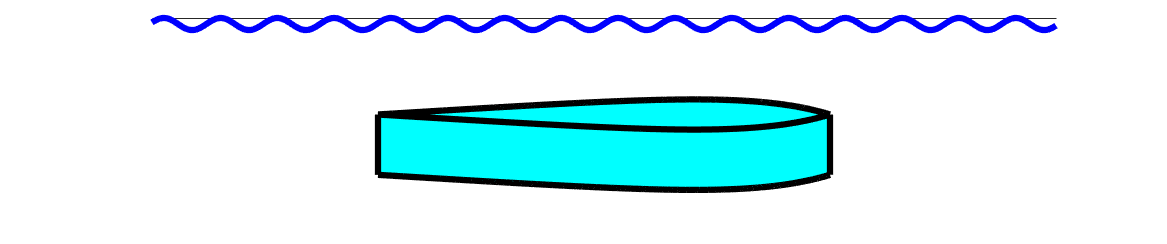}}
\end{overpic}};
\draw[line width=1,->] (2.6,4.3) -- (2.6,4.8);
\draw[line width=1,->] (2.6,6.8) -- (2.6,6.3);
\draw[line width=0.5] (2.3,4.8) -- (3.0,4.8);
\draw[line width=1,->] (-2.6,4.3) -- (-2.6,4.8);
\draw[line width=1,->] (-2.6,5.9) -- (-2.6,5.4);
\draw[line width=0.5] (-2.1,4.8) -- (-3.0,4.8);
\draw[line width=0.5] (-2.1,5.4) -- (-3.0,5.4);
\node at (-3.5,5.0){$H$};
\node at (2.6,5.5){$D$};
\node at (0,7.2) {\bf Depth $\boldsymbol{d=D/H=2.0}$};
\end{tikzpicture}
\caption{Experimentally measured drag coefficient (\ref{Cdeq}) for hull 5 from the slender family at different Froude numbers and at depths (a) $d=0.5$ and (b) $d=2.0$. The difference between positive and negative asymmetry $\epsilon$ is indicated. Error bars correspond to one standard deviation of the time signal given by the force sensor. \label{2depths}}
\end{figure}

In figure \ref{2depths} we display drag coefficients measured for hull 5 from the slender family at depth $d=0.5$ (a) and $d=2.0$ (b) at Froude numbers between $0.3$ and $1.0$. For the shallow case $d=0.5$, the drag coefficient for $\epsilon<0$ is lower than $\epsilon>0$ for all measured Froude numbers. 
This is because when the body moves at the water surface, the total drag is dominated by its wave drag component, and
for $\epsilon>0$, where the leading edge is less pointed than the trailing edge, there is a larger wave disturbance than for $\epsilon<0$. 
For the deep case $d=2.0$ (figure \ref{2depths}(b)), the total drag is dominated by its form drag component. Hence, for all Froude numbers there is lower drag when the body moves in its more streamlined direction, with its more pointed end at the trailing edge ($\epsilon>0$).

In the intermediate depths between $d=0.5$ and $2.0$, as we will discuss later, neither $\epsilon<0$ nor $\epsilon>0$ is optimal for all Froude numbers. Instead, there is a range of Froude numbers for which wave drag is more important than form drag (and $\epsilon<0$ optimal) and the complimentary range where form drag is more important (and $\epsilon>0$ optimal).

%\clearpage

\begin{figure}
\centering
%\vspace{0.5cm}
\begin{overpic}[width=0.15\textwidth]{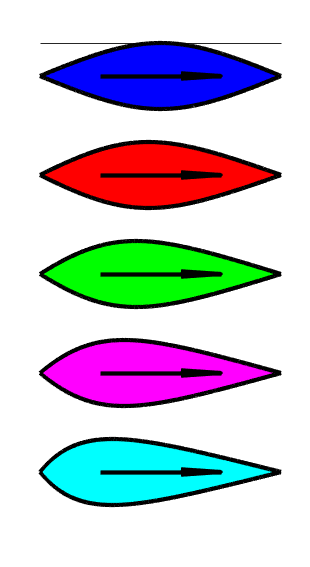}
\put  (0,100) {\textit{Colour scheme}}
\put  (45,115) {(a)}
\end{overpic}
\begin{overpic}[width=0.68\textwidth]{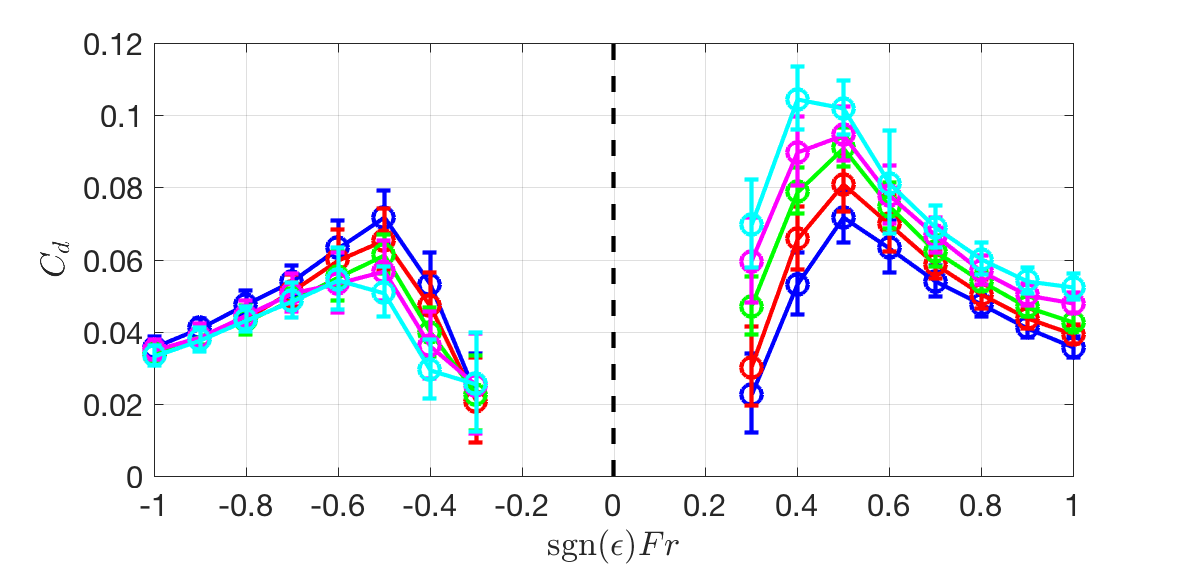}
\put  (32,47) {\bf Tow-tank measurements}
\put(72,25){\color{black}\vector(0,1){18}}
\put(32,35){\color{black}\vector(0,-1){18}}
%\put  (60,12) {\includegraphics[width=0.17\textwidth]{figs/analytical_nbl}}
%\put  (20,14) {\bf Michell theory}
%\put(48,14){\color{black}\vector(1,0){10}}
\end{overpic}
\begin{overpic}[width=0.15\textwidth]{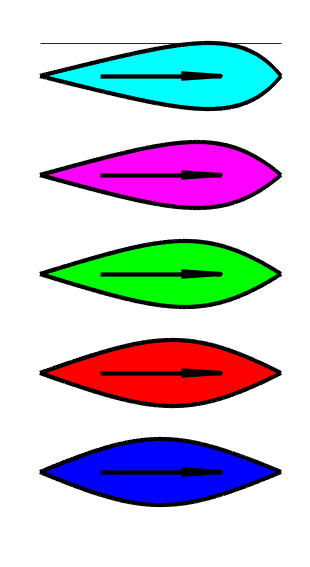}
\put  (5,100) {\textit{Colour scheme}}
\end{overpic}\\
\vspace{0.4cm}
\begin{overpic}[width=0.47\textwidth]{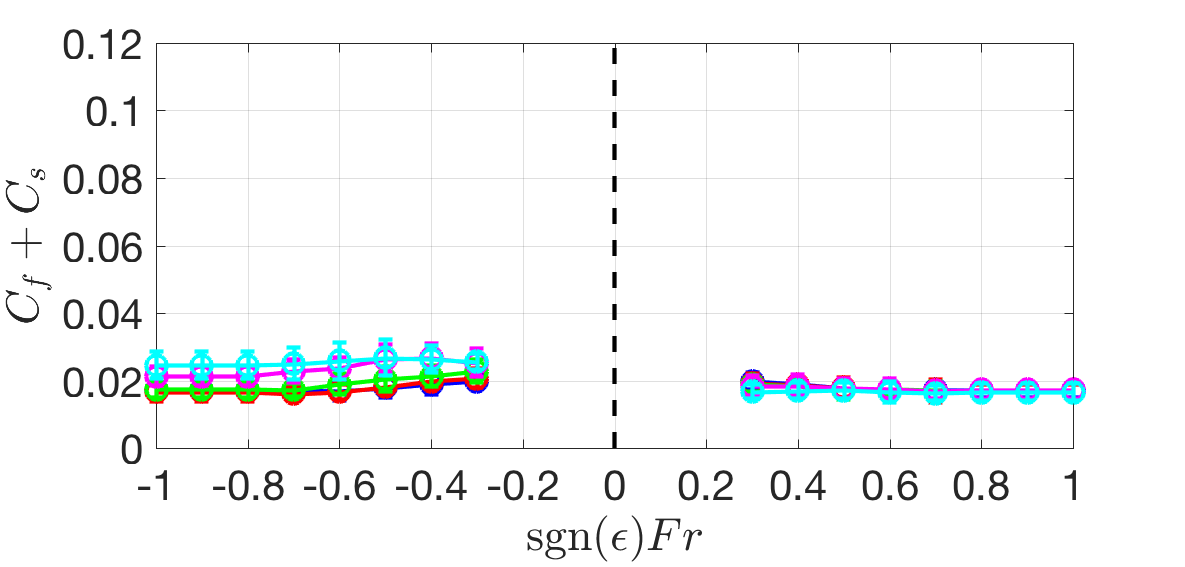}
\put  (15,47) {\bf Wind tunnel measurements}
\put  (0,47) {(b)}
\put(25,11){\color{black}\vector(0,1){12}}
\put(65,19){\color{black}\vector(0,-1){8}}
%\put(52,30){\color{black}\vector(-1,0){10}}
%\put (44,32) {\textit{Symmetric}}
\end{overpic}
\begin{overpic}[width=0.47\textwidth]{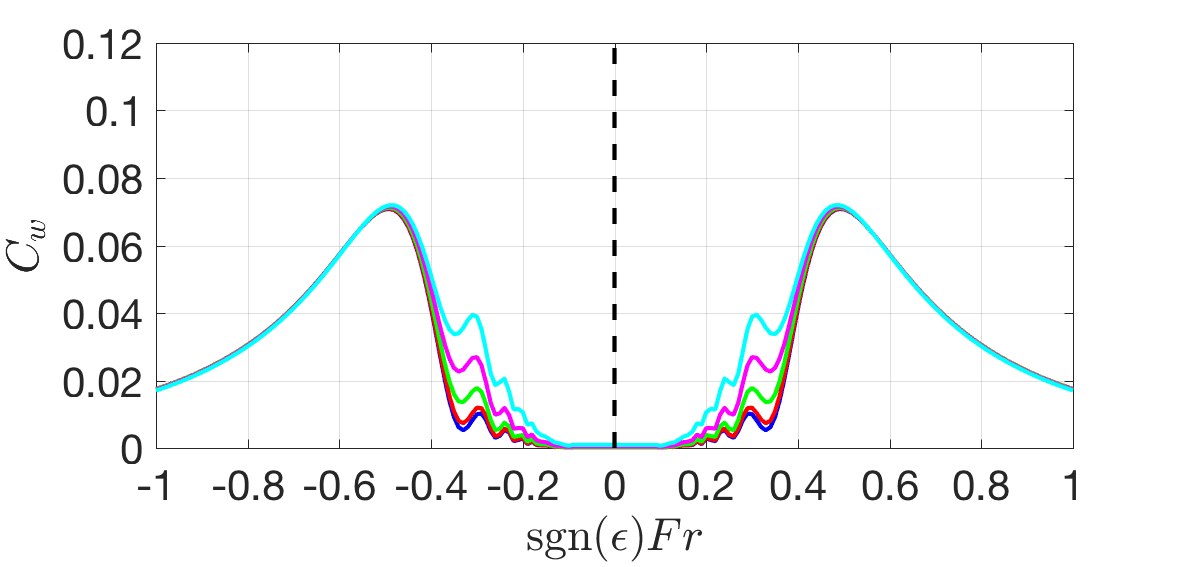}
\put  (30,47) {\bf Michell's theory}
\put  (0,47) {(c)}
%\put(52,30){\color{black}\vector(1,0){10}}
%\put(52,30){\color{black}\vector(-1,0){10}}
%\put (44,32) {\textit{Symmetric}}
\end{overpic}
\caption{(a) Drag coefficient (\ref{Cdeq}) measured using a tow-tank experiment for each of the 5 hulls from the slender family with both positive and negative $\epsilon$ at fixed depth $d=0.5$ and at different Froude numbers. (b) Form and skin drag coefficient $C_f+C_s$ (\ref{windskin}) measured in a wind tunnel, in absence of the air-water interface. (c) Theoretical \textit{symmetric} prediction of the wave drag coefficient $C_w$ using Michell's formula (\ref{Cweq}). \label{exp}}
\end{figure}
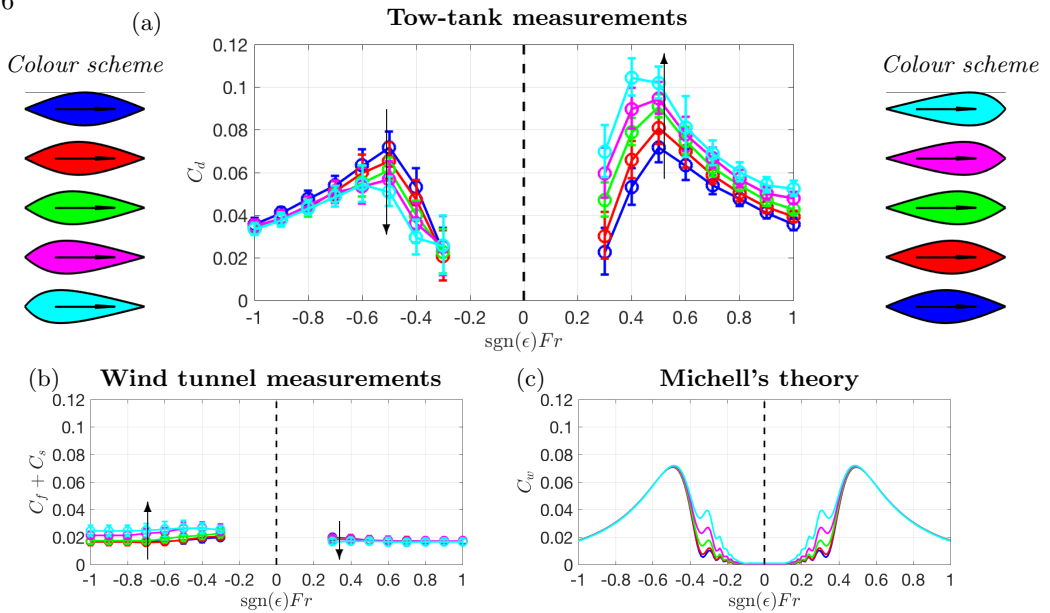

Next we investigate the other hull shapes from the same slender family. In figure \ref{exp} we show measured values of $C_d$ for $d=0.5$, for each of the $5$ different hull shapes, and in both directions of motion. We can see that the drag increases with increasing $\epsilon>0$ (i.e. for hull number from 1 to 5), whereas drag decreases with decreasing $\epsilon<0$. This is consistent with figure \ref{2depths}, and indicates that bodies with pointed leading edges are optimal at shallow depths. The drag curves exhibit characteristic maxima near $\mathrm{\it Fr}= 0.5$, as is often seen in the literature \citep{tuck1989wave,videler2012fish}.

There are three major contributions to the measured drag: wave drag, form drag and skin drag \citep{newman2018marine}. 
We have already described wave and form drag, and skin drag is the force associated with viscous friction on the wetted surface of the hull. We write the total drag coefficient\footnote{Note that the skin and form drag, $C_s$ and $C_f$, are classically normalised by a factor $1/2\rho U^2 S$, where $S$ is the wetted surface area. Here, they form part of the total drag coefficient $C_d$, and so are normalised by $\rho U^2 \Omega^{2/3}$, as in (\ref{Cdeq}).} in terms of this decomposition  
\beq
C_d=C_w+C_f+C_s.\label{allCs}
\eeq
It is difficult to isolate and measure any one of these components. However, by measuring the drag on the hull shapes when placed in a wind tunnel, it is possible to isolate the skin and form contributions. 
In doing this, we make the key assumption that the air-water interface does not affect the skin and form drag. For the small amplitude waves we consider, we expect this assumption to be valid since the problem is sufficiently linear that the coupling is weak.

%In Figure \ref{exp} (b) we display the drag coefficients measured using the wind tunnel. 
To compare the wind tunnel and the tow-tank measurements, we use equivalent values of the Reynolds number in both air and water. For the tow-tank experiment the Reynolds number is given in terms of the Froude number by $\Rey_{\mathrm{water}}=\mathrm{\it Fr}L\sqrt{gL}/\nu_{\mathrm{water}}$, where $\nu$ denotes the kinematic viscosity.
The Reynolds number of the air in the wind tunnel is given by $\Rey_{\mathrm{air}}=U_{\mathrm{air}} L/\nu_{\mathrm{air}}$. 
By equating these, we obtain the required air velocity, which is given in terms of the Froude number as $U_{\mathrm{air}}=\mathrm{\it Fr} \sqrt{g L} (\nu_{\mathrm{air}}/\nu_{\mathrm{water}})$.

In figure \ref{exp}(b) we display measurements of the combined skin and form drag from the wind tunnel experiments, where the measured force $R_{\mathrm{air}}$ is converted into the sum of the drag coefficients by the relationship
\beq
C_f+C_s=\frac{R_{\mathrm{air}}}{\rho_{\mathrm{air}} U_{\mathrm{air}}^2\Omega^{2/3}}.\label{windskin}
\eeq
The measured coefficients in the plot are given in terms of the Froude number for the purpose of comparison with (a). In contrast to the tow-tank experiments, here we see an increase in drag for decreasing $\epsilon<0$, and a decrease in drag for increasing $\epsilon>0$. 
This is expected since hull shapes with $\epsilon>0$ are more streamlined for hull numbers increasing from 1 to 5, whereas hull shapes with $\epsilon<0$ are less streamlined.
Hence, it is clear that the effect of asymmetry on skin and form drag is completely the opposite as on wave drag.

We also observe that the contribution of form and skin drag in this case is relatively small compared to the total drag. Therefore, it is expected that the wave drag component is responsible for the dominant behaviour seen in (a). In particular, the characteristic shape of the drag curves in (a), with peaks near $\mathrm{\it Fr}= 0.5$ are typical of wave drag measurements \citep{tuck1989wave,videler2012fish}.

Now let us consider Michell's theoretical prediction of the wave drag $C_w$ \citep{michell1898xi,tuck1989wave}, which is given by the formula
\beq
C_w = \frac{4\beta^{2/3}}{\pi \alpha^{4/3}\mathrm{\it Fr}^4}\int_1^\infty \frac{\lambda^2}{\sqrt{\lambda^2-1}}|G(\lambda)|^2 \,\mathrm{d}\lambda,\label{Cweq}
\eeq
where $\alpha=L/W$ and $\beta=L/H_w$ are the horizontal and vertical aspect ratios of the wetted hull, and $G$ is an integral expression defined in terms of non-dimensional variables as
\beq
%G= \frac{1}{HL^2} \int_0^\infty\int_{-\infty}^{\infty} \left[\boldsymbol{f(x)} \right]e^{(-\lambda^2 z/L+i\lambda x/L)/Fr^2}\,\mathrm{d}{x}\,\mathrm{d}{z}.\label{CwG}
G(\lambda)=  \int_{-D/L}^{(-D+H_w)/L}\int_{-1/2}^{1/2} \left[ {\hat{f}'(\hat{x})} \right] e^{(-\lambda^2 \hat{z}+i\lambda \hat{x})/\mathrm{\it Fr}^2}\,\mathrm{d}{\hat{x}}\,\mathrm{d}{\hat{z}}.\label{CwG}
\eeq
The central term in (\ref{CwG}), $\hat{f}'(\hat{x})$, corresponds to a distribution of sources located along the $\hat{x}$ axis, with strength equivalent to forcing the impermeability condition along the body walls $\hat{y}=\pm \hat{f}(\hat{x})$. Clearly, if $\hat{f}(\hat{x})$ is an asymmetric function, it makes no difference to (\ref{CwG}) whether the body moves forwards or backwards (consider the transformation $\hat{x}\rightarrow -\hat{x}$).

We use (\ref{Cweq})-(\ref{CwG}) to compute the theoretical prediction of $C_w$, which we display in figure \ref{exp}(c). 
We can see that Michell's formula captures the general behaviour of the wave drag, exhibiting the characteristic peaks near $\mathrm{\it Fr}= 0.5$, as seen in (a). However, it fails to distinguish between forward and backward motion. 
Indeed, if we sum together Michell's wave drag prediction $C_w$ with the wind tunnel measurements of form and skin drag $C_f+C_s$, the only asymmetry effect observed comes from the form and skin components, which display the opposite trend to the tow-tank experiments in (a).
%Hence, Michell's theory is not at all suitable here.
Furthermore, apart from at small Froude numbers, there is no significant difference in the drag between the 5 hull shapes. 
Hence, such a model cannot be used to replicate the observed experimental results, and cannot be used to accurately search design spaces, or to find optimum asymmetry, for example. 

%\clearpage

We have also performed experimental measurements for each of the 5 slender and bluff bodies at depths between $d=0.5$ and $2.0$ and at Froude numbers between $0.3$ and $1.5$, and these are presented in Appendix \ref{appB}.

%\clearpage

\section{Breaking the symmetry}

In this section we use a variety of theoretical approaches to interpret the asymmetry effects observed in our experimental results. We start by proving that the wave resistance problem, as formulated by Michell using the steady Euler equations, has an inherent symmetry, rendering it incapable of predicting asymmetry effects. By adding dissipation the symmetry is broken. Hence, by using a $k$-$\omega$ SST model (which is dissipative) we show that our experimentally observed asymmetry effects can be replicated. Finally, as a simpler alternative approach, we show that these asymmetry effects can also be captured by modifying Michell's theory to account for the growth of a turbulent boundary layer.

\subsection{A note on reversibility}\label{reverse}

First, we describe the original formulation of the problem described by \citet{michell1898xi}. We revert back to dimensional coordinates $(x,y,z)$ for convenience. 
In this framework, the velocity and pressure are denoted $\boldsymbol{u}=(u,v,w)$ and $p$.
Assuming incompressible, inviscid flow, the governing equations are the steady Euler equations 
\begin{align}
\nabla \cdot \boldsymbol{u}&=0,\label{euler1}\\
\rho \lb \boldsymbol{u}\cdot \nabla \rb \boldsymbol{u} &=-\nabla p - \rho g \hat{\boldsymbol{k}},\label{euler2}
\end{align}
where $\hat{\boldsymbol{k}}$ is the unit vector in the vertical $z$ direction. The boundary conditions consist of the impermeability conditions on the hull walls, which are
\beq
v=\pm u f'(x),\quad\mathrm{on}\quad y=\pm f(x),\label{imperm}
\eeq
the kinematic and dynamic conditions at the air-water interface $z=\zeta(x,y)$, which are
\begin{align}
w=u \zeta_x + v \zeta_y,\quad &\mathrm{on}\quad z=\zeta(x,y),\\
p=p_{atm},\quad &\mathrm{on}\quad z=\zeta(x,y),
\end{align}
as well as appropriate conditions at infinity
\beq
\boldsymbol{u}\rightarrow (U,0,0),\quad x,y,z\rightarrow\pm\infty.\label{decay}
\eeq
Following this, Michell then assumes an irrotational flow so that the above formulation can be written in terms of a velocity potential, and then applies slender body theory to linearise the boundary conditions. However, it is clear that, even before making these final assumptions, there is already an inherent symmetry in the problem formulation.

To illustrate this, first consider that $\boldsymbol{u}^*$, $p^*$ and $\zeta^*$ are solutions to the free boundary problem (\ref{euler1})-(\ref{decay}). 
Then consider switching the direction of the free stream $U\rightarrow -U$. It is straightforward to show that the reversed flow problem has a solution $-\boldsymbol{u}^*$, $p^*$ and $\zeta^*$, regardless of whether $\hat{f}(\hat{x})$ is an asymmetric function.
Hence, the problem is invariant under a change in the direction of motion. Consequently, such a formulation cannot predict the effects of asymmetry, such as those we have observed experimentally.

%The reversibility of the equations derives from having neglected viscosity. 
There are several possible explanations for the failure of the above formulation to capture asymmetry effects.
Firstly, we observe that by including a viscous term $\mu \nabla^2 \boldsymbol{u}$ on the right hand side of (\ref{euler2}), the variables $-\boldsymbol{u}^*$, $p^*$ and $\zeta^*$ no longer satisfy the reversed flow problem. This indicates that neglecting viscosity in the Euler equations may be responsible for the failure. 
One could similarly argue that the failure is caused by choosing the steady equations, which neglect the acceleration term $\rho \partial \boldsymbol{u}/\partial t$ term on the left hand side of (\ref{euler2}).

Here, we show that it is sufficient to account for viscosity to capture the effect of body asymmetry on drag. We illustrate this in two ways. First, with a steady $k$-$\omega$ SST turbulence model. Then, by modifying Michell's theory to include the growth of a turbulent boundary layer.

\subsection{Results from a $k$-$\omega$ SST model}

Since we consider situations where the Reynolds number is between $\Rey=10^5$ and $\Rey=10^8$, the flow near the hull is expected to be turbulent. Hence, we model the flow with a steady three-dimensional $k$-$\omega$ SST model \citep{menter1992improved}, where the air-water interface is treated with the volume of fluid method \citep{ubbink1997numerical,berberovic2009drop}. The momentum equation of the $k$-$\omega$ SST model contains a term on the right hand side of the form $(\nabla{:}(\mu+\mu_t)(\nabla \boldsymbol{u}+ \nabla\boldsymbol{u}^T))$, where $\mu_t=\mu_t(k,\omega)$ is a non-linear eddy viscosity. As explained above, this term breaks the symmetry of the problem, allowing us to distinguish between forward and backward motion.

Since it is not possible to perform computations on an infinite domain, instead we use a finite domain with boundaries more than $20$ hull lengths away from the centre of the hull. We find this is sufficient to avoid significant effects due to wave reflections from the edges. In addition, since the problem is symmetric about the plane $y=0$, we only solve for $y\geq 0$. We use a cuboid mesh with $(40,20,120)$ elements in the $(x,y,z)$ directions, spaced non-uniformly such that the resolution near the hull walls and at the air-water interface is much higher than in the far field. We have also tried finer mesh resolutions, and we find that this mesh resolution is sufficient to resolve all the details of the flow.

In addition to the boundary conditions (\ref{imperm})-(\ref{decay}), we also impose no-slip conditions on the hull walls, and appropriate conditions for the turbulence variables $k$ and $\omega$, which we do not describe here, but which are given by \citet{menter1992improved}. We use standard values for all the turbulence parameters, which are also given by \citet{menter1992improved}.

\begin{figure}
\centering
\begin{overpic}[width=0.49\textwidth]{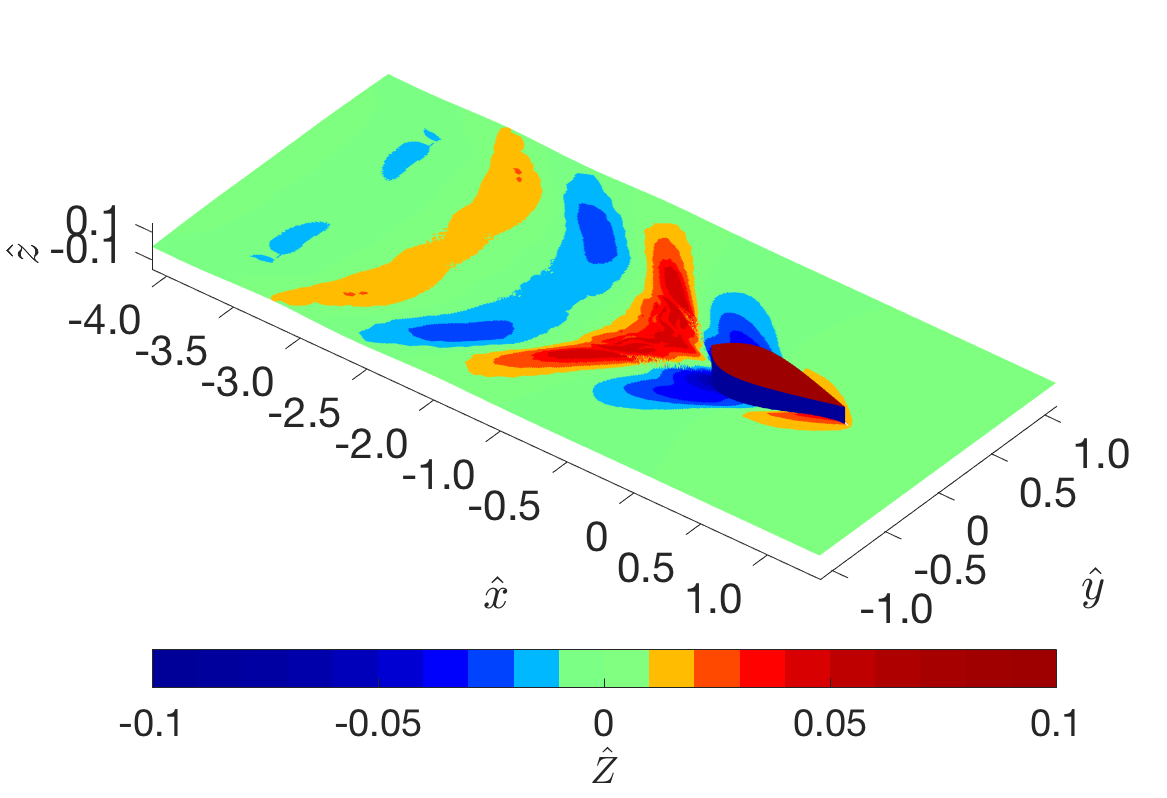}
\put (50,55) { $\boldsymbol{\epsilon<0}$}
\put (5,55) {(a)}
\end{overpic}
\begin{overpic}[width=0.49\textwidth]{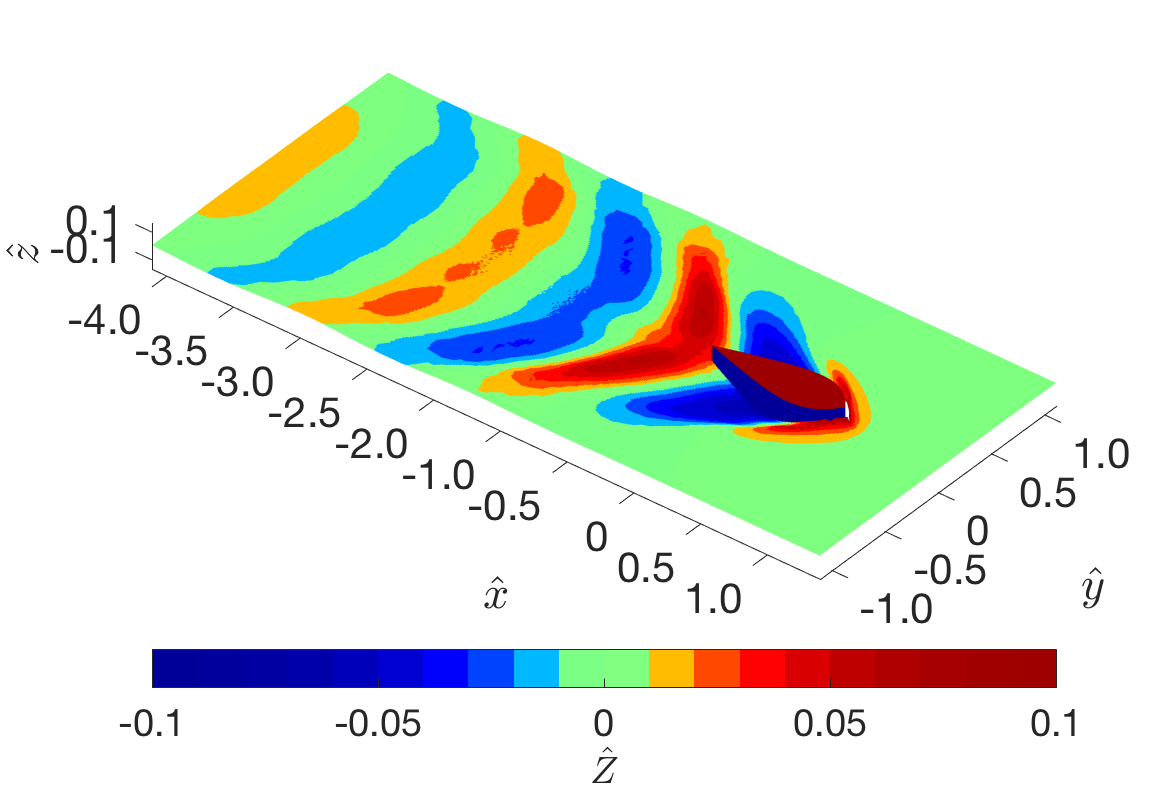}
\put (52,55) { $\boldsymbol{\epsilon>0}$}
\put  (-40,65) {\bf Numerical results ($k-\omega$ SST)}
\put (5,55) {(b)}
\end{overpic}\\
\vspace{0.2cm}
\begin{overpic}[width=0.15\textwidth]{figs/all_logs_b}
\put  (0,100) {\textit{Colour scheme}}
\put (10,115) {(c)}
\end{overpic}
\begin{overpic}[width=0.68\textwidth]{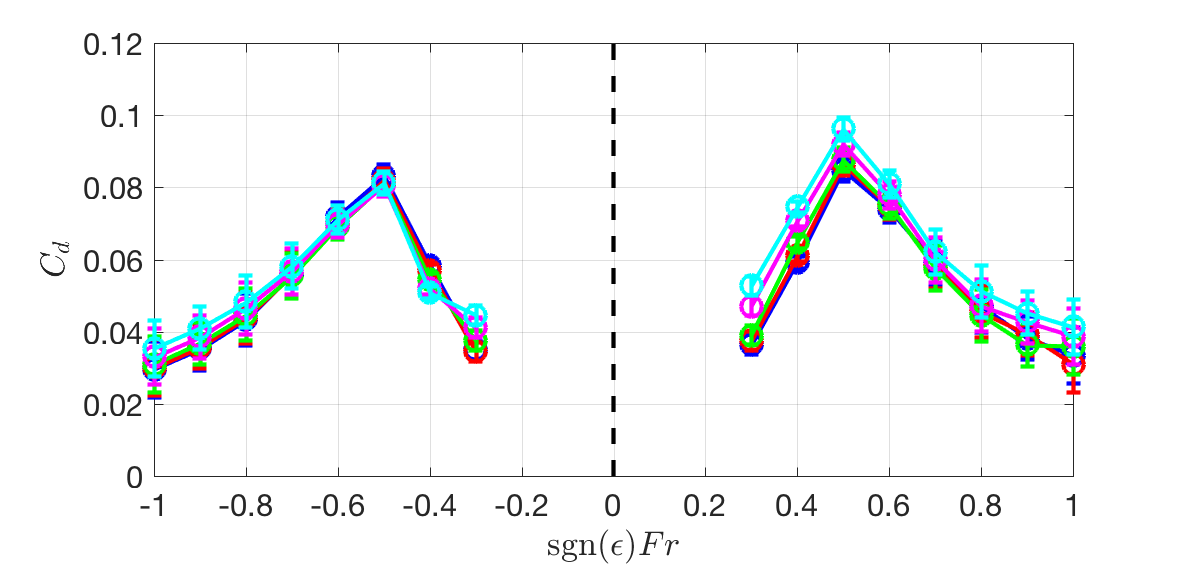}
%\put  (32,47) {\bf Experimental results}
%\put  (60,12) {\includegraphics[width=0.17\textwidth]{figs/analytical_nbl}}
%\put  (20,14) {\bf Michell theory}
\put(72,30){\color{black}\vector(0,1){10}}
\put(32,38){\color{black}\vector(0,-1){10}}
\end{overpic}
\begin{overpic}[width=0.15\textwidth]{figs/all_logs_f}
\put  (5,100) {\textit{Colour scheme}}
\end{overpic}
%\begin{overpic}[width=0.64\textwidth]{figs/numerics}
%\put(55,40){\color{black}\vector(1,3){12}}
%\put (-0,45) {(c)}
%\end{overpic}
\caption{Computational results from a $k$-$\omega$ SST model at depth $d=0.5$. (a, b) Surface plots of the air-water interface for hull 5 from the slender family at Froude number $\mathrm{\it Fr}=0.5$ with positive and negative $\epsilon$. (c) Numerically computed drag coefficient for comparison with figure \ref{exp}.  \label{numerics}}
\end{figure}
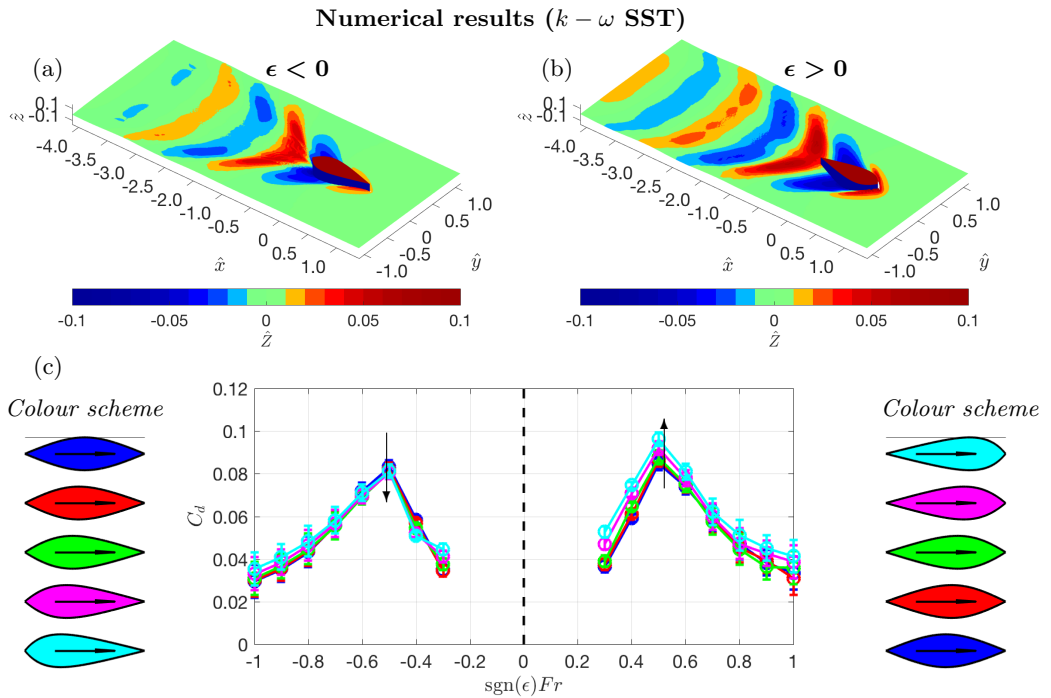

In figure \ref{numerics}(a,b) we display surface plots of the air-water interface solution for the case of hull 5 (from the slender family) with both positive and negative $\epsilon$. In each case the Froude number is $\mathrm{\it Fr}=0.5$. We can see the classic Kelvin wake pattern behind the hulls, though the waves for $\epsilon>0$ are larger in amplitude and persist further downstream, illustrating just one asymmetry effect.
 %Hence, clearly the $k$-$\omega$ SST model is capable of capturing asymmetry effects. 

In figure \ref{numerics}(c) we display drag coefficients calculated for all 5 hull shapes from the slender family, with both positive and negative $\epsilon$, and for Froude numbers $\mathrm{\it Fr}=0.3-1.0$. The depth is fixed at $d=0.5$ for the sake of comparison with figure \ref{exp}. In each case the drag coefficients are calculated by integrating the stress around the hull surface and normalising by a factor $\rho U^2 \Omega^{2/3}$, as in (\ref{Cdeq}). 

Overall, there is relatively good comparison between the experimental and numerical results. The $k$-$\omega$ SST model captures the correct magnitude of the drag coefficient, as well as the appropriate increase in $C_d$ for increasing $\epsilon>0$ and the decrease for decreasing $\epsilon<0$. However, the difference in $C_d$ between the different hull shapes is not as large as measured in the experiments. This discrepancy could possibly be due to inaccurate treatment of the air-water interface, or the development of the turbulent boundary layer. The discrepancy might be resolved by using a LES or DNS simulation instead of a RANS model, though this would be significantly more computationally intensive.
Nevertheless, the $k$-$\omega$ SST model is clearly capable of capturing asymmetry effects, at least qualitatively. 

We have also performed computations for the bluff family of shapes at various depths, and these are presented in Appendix \ref{appB}. In particular, the bluff family of shapes exhibit more of an extreme difference in drag between $\epsilon>0$ and $\epsilon<0$, and this is detected more clearly with the $k$-$\omega$ SST simulations (e.g. figure \ref{app_bluff}(e)).

\subsection{Modification of Michell's theory}

Next we show that a simple modification to Michell's theory can account for 
the effects of body asymmetry, capturing the distinction between forward and backward motion.  
%In our approach, we modify the formula for the wave drag (\ref{CwG}) to account for the growth of a boundary layer. 
A possible theoretical underpinning for such distinction is the effect of viscosity. 
Since, with a viscous description of the flow, there is a boundary layer near the hull walls, our approach here is to modify Michell's theory to account for the growth of this boundary layer. 
We find that this effective approach captures the trends observed experimentally.

With a viscous description of the flow, the impermeability condition (\ref{imperm}) is replaced by a no-slip condition on the body walls. The flow is then decomposed into an inner boundary layer region, where the effect of viscosity is important, and an outer inviscid potential-flow region. 
In our current approach, we treat the edge of the boundary layer as an impermeable surface to the potential flow region (or equivalently a streamline which passes around the hull). 
Then, we advance in the same manner as \citet{michell1898xi}, as described in Section \ref{reverse}, except we impose the impermeability condition (\ref{imperm}) on the combined shape of the hull plus its boundary layer. 
By doing so, we replace the hull with a new larger shape which has non-zero width at the trailing edge. 
We expect the boundary layer to have approximately the same aspect ratio as the hull shape, so that slender body theory still applies. 
Furthermore, the wave drag force due to pressure variations along the hull wall are transmitted to the edge of the boundary layer, since pressure is expected to be uniform across the boundary layer width \citep{schlichting1960boundary}.

For the purposes of this study, we take the boundary layer thickness as the $99\%$ definition $\delta_{0.99}(x)$: for a given $(x, y)$ plane, this is defined as the $y$ value that corresponds to where the streamwise velocity is at $99\%$ of its maximum $u(x,y=\delta_{0.99}(x))=0.99\max \{u(x,y)\}$. Inserting the combined shape of the hull plus its boundary layer into Michell's non-dimensional formula for the wave drag (\ref{CwG}), we get
\beq
G(\lambda)=  \int_{-D/L}^{(-D+H_w)/L}\int_{-1/2}^{1/2} \left[ {\hat{f}'(\hat{x})+\hat{\delta}_{0.99}'(\hat{x}) } \right] e^{(-\lambda^2 \hat{z}+i\lambda \hat{x})/\mathrm{\it Fr}^2}\,\mathrm{d}{\hat{x}}\,\mathrm{d}{\hat{z}}.\label{CwG2}
\eeq
Here, we have made the assumption that $\hat{\delta}_{0.99}'(\hat{x})=0$ for $|\hat{x}|>1/2$. 
This is equivalent to a boundary layer which begins growing at the leading edge and, at the trailing edge, it turns into a wake region which remains at constant width downstream.

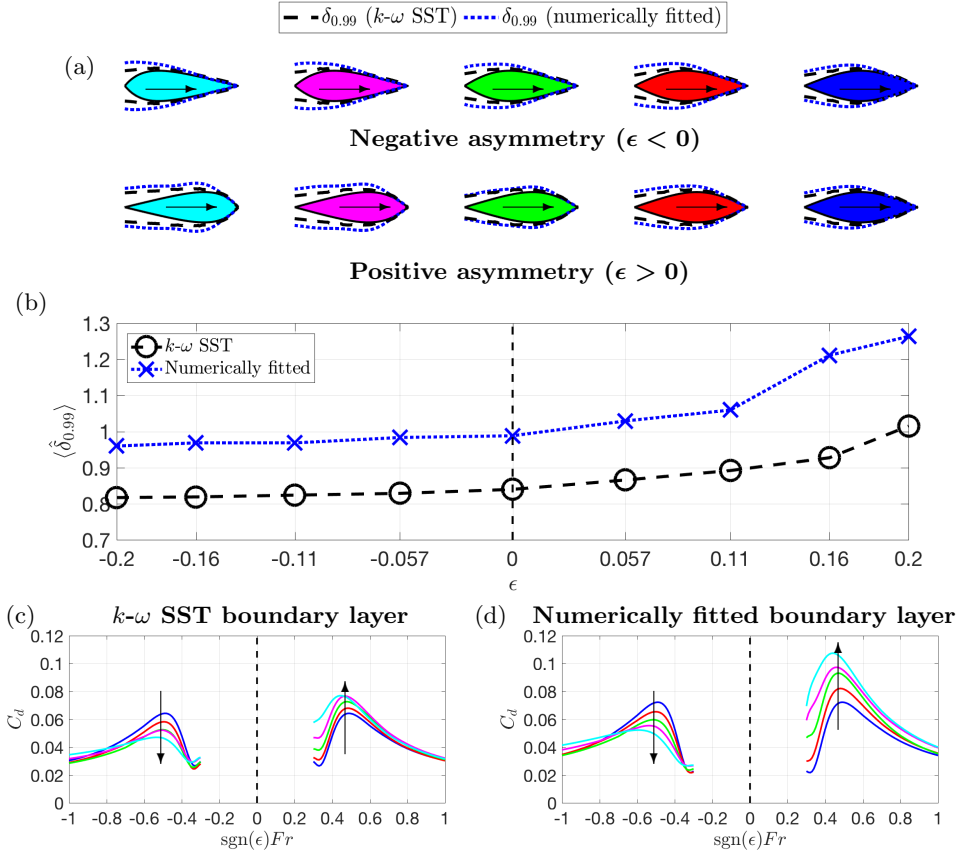
\begin{figure}
\centering
\begin{tikzpicture}[scale=0.4]
\node at (0,0) {\begin{overpic}[width=1\textwidth]{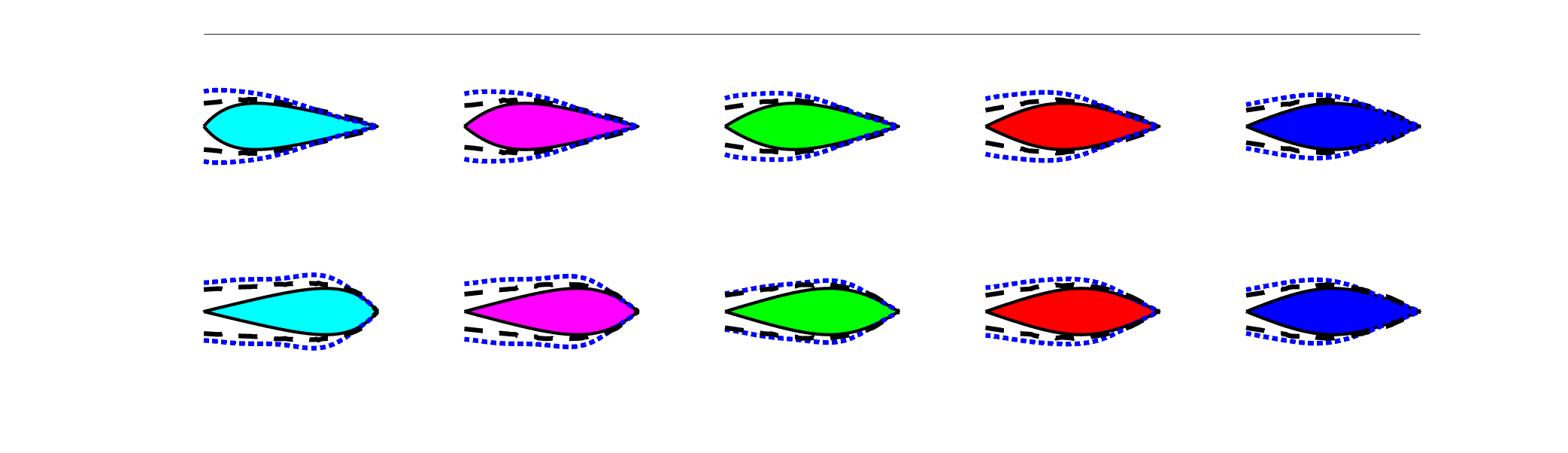}
\put (35,15) {\bf Negative asymmetry ($\boldsymbol{\epsilon<0}$)}
\put (7,22) {(a)}
%\put(40,26){\color{blue}\vector(-2,-1){5}}
%\put (41, 26) {\bf \textit{\color{blue}Effective boundary layer}}
\put(83,20.5){\color{black}\vector(1,0){5}}
\put(66,20.5){\color{black}\vector(1,0){5}}
\put(49,20.5){\color{black}\vector(1,0){5}}
\put(32,20.5){\color{black}\vector(1,0){5}}
\put(15,20.5){\color{black}\vector(1,0){5}}
\put (35,2) {\bf Positive asymmetry ($\boldsymbol{\epsilon>0}$)}
\put(83,9){\color{black}\vector(1,0){5}}
\put(67,9){\color{black}\vector(1,0){5}}
\put(50,9){\color{black}\vector(1,0){5}}
\put(34,9){\color{black}\vector(1,0){5}}
\put(17,9){\color{black}\vector(1,0){5}}
\end{overpic}};
\draw[line width=0.0, white, fill=white] (-15,4) rectangle (15,5.0);
%\node at (0,4.5) {\bf \textit{\color{blue}Effective boundary layer}};
\node at (0,4.5) {\includegraphics[width=0.45\textwidth]{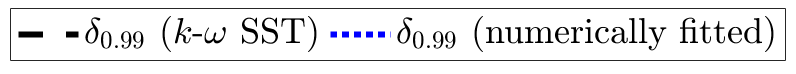}};
\end{tikzpicture}\\
%\vspace{0.2cm}
\begin{overpic}[width=1\textwidth]{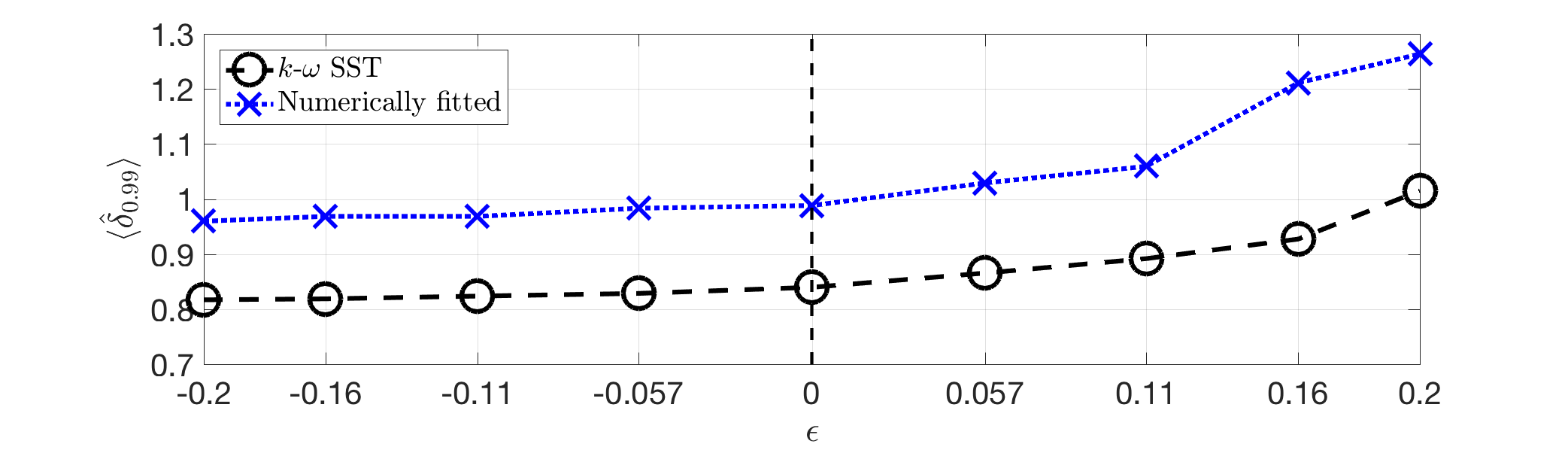}
\put (3,28) {(b)}
\end{overpic}\\
\vspace{0.3cm}
\begin{overpic}[width=0.475\textwidth]{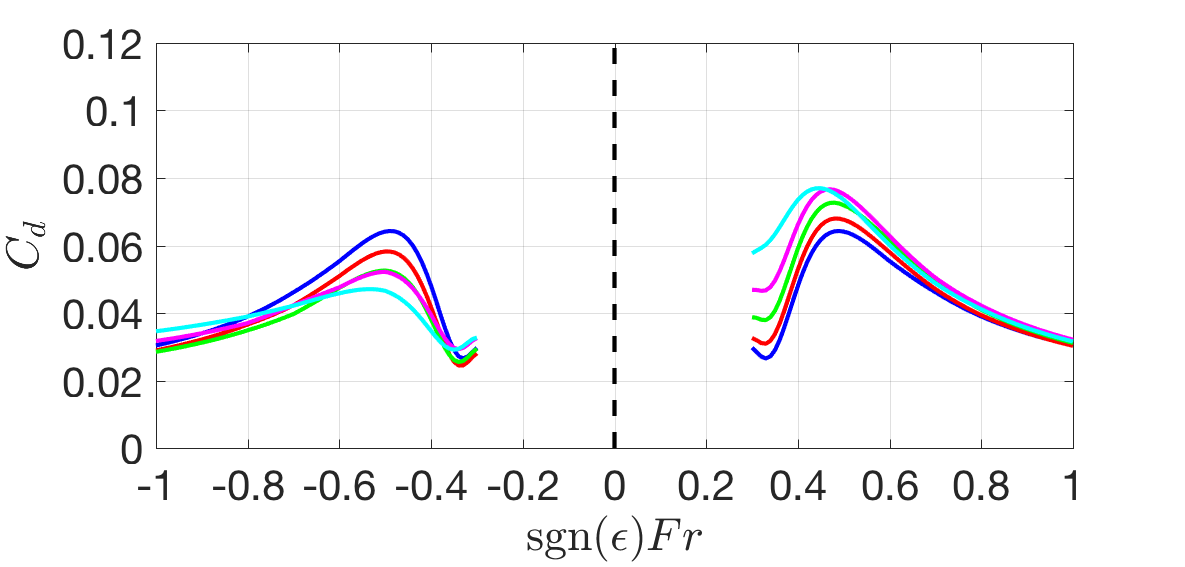}
\put  (22,47) {\bf $k$-$\omega$ SST boundary layer}
\put (0,47) {(c)}
\put(70,20){\color{black}\vector(0,1){15}}
\put(32,33){\color{black}\vector(0,-1){15}}
\end{overpic}
\begin{overpic}[width=0.475\textwidth]{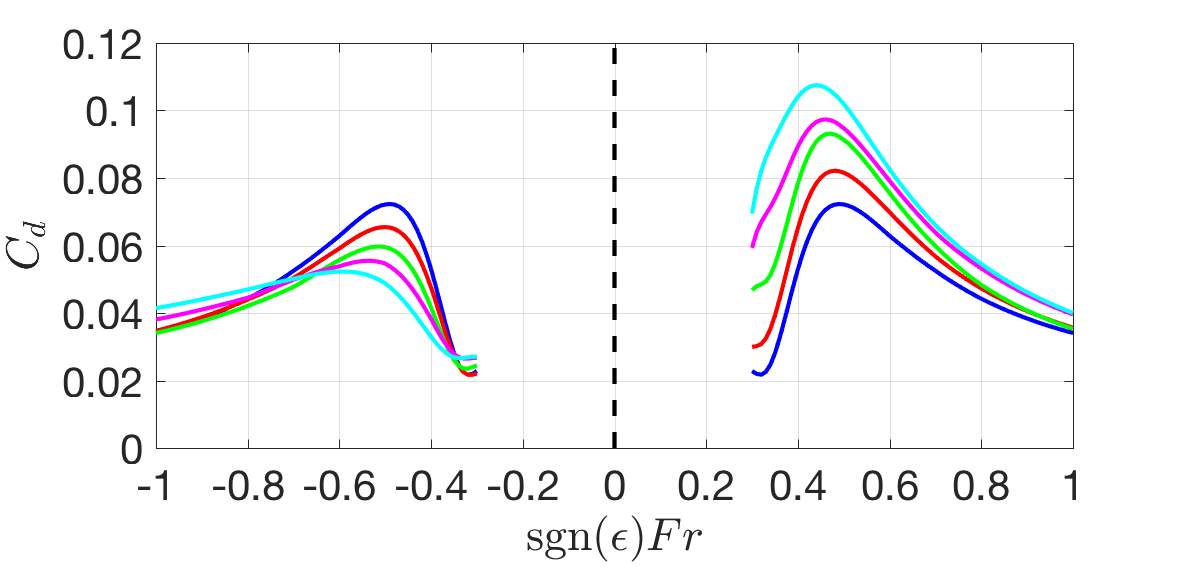}
\put  (7,47) {\bf Numerically fitted boundary layer}
\put (-5,47) {(d)}
\put(70,25){\color{black}\vector(0,1){18}}
\put(32,33){\color{black}\vector(0,-1){15}}
\end{overpic}
\caption{(a) Turbulent boundary layers for the five hulls from the slender family, with $\epsilon>0$ and $\epsilon<0$. For the $k$-$\omega$ SST model we take the boundary layer as the $99\%$ thickness. For the numerically fitted boundary layer, we use (\ref{CwG2}) and a numerical optimisation routine to find the boundary layer profile that best matches the theoretical and experimental drag coefficients. (b) Corresponding average boundary layer thickness. (c, d) Theoretical drag coefficients calculated using (\ref{CwG2}) with the corresponding boundary layer profiles, in conjunction with wind tunnel measurements for $C_f+C_s$ (to be compared with figures \ref{exp} and \ref{numerics}).\label{anal}}
\end{figure}

To estimate the boundary layer thickness, we make use of our $k$-$\omega$ SST simulations. 
To reduce noise, we extract the boundary layer thickness from simulations where the hull is deeply submerged beneath the water surface ($d=2.0$).
We assume that the boundary layer profile does not change much with depth, and it is therefore acceptable to use the profile measured at $d=2.0$ for all depths.
% where we define the boundary layer thickness at a given $x$ value as $u(x,y=\delta_{0.99}(x))=0.99\max u(x,y)$. 
The boundary layer thickness for each hull shape is displayed in figure \ref{anal}(a) with black dashed lines. For $\epsilon<0$ the boundary layer grows slowly for the majority of the hull shape, and then very rapidly at the trailing edge. By contrast, for $\epsilon>0$ the boundary layer only grows slowly near the leading edge, and then rapidly thereafter. 
A useful measure for the size of the boundary layer is the average width $\langle\hat{\delta}_{0.99}\rangle=\int_{-1/2}^{1/2}\hat{\delta}_{0.99}\,d\hat{x}$.
We plot $\langle\hat{\delta}_{0.99}\rangle$ calculated for each of the hull shapes in (b). Clearly, we see that the average boundary layer thickness is larger for $\epsilon>0$ than for $\epsilon<0$ (e.g. $24\%$ larger for hull 5).

Using the modification to Michell's theory (\ref{CwG2}) with the boundary layer thicknesses extracted from the $k$-$\omega$ simulations, in conjunction with the wind tunnel measurements of the skin and form drag, we calculate the total drag (\ref{allCs}) for each of the 5 slender hulls with both $\epsilon<0$ and $\epsilon>0$. The results are plotted in figure \ref{anal}(c). 
We see that the trend observed in the experimental results from figure \ref{exp} is replicated very well, even better than the numerical calculations in figure \ref{numerics}. 
The hulls with positive $\epsilon$ have increased drag, whilst those with negative $\epsilon$ have decreased drag. 
However, there is clearly still some discrepancy for the shapes with large $\epsilon>0$ (e.g. the drag on hull 5 at $\mathrm{\it Fr}=0.5$ is too small).

It is interesting to note that even though the combined shape of the hull plus its boundary layer is bigger than the original hull $\hat{f}+\hat{\delta}_{0.99}\geq\hat{f}$, the modification  (\ref{CwG2}) can produce either an increase or a decrease in wave drag, depending on the sign of $\epsilon$. Therefore, the average width $\langle\hat{\delta}_{0.99}\rangle$ does not provide enough information alone to indicate whether the asymmetry is advantageous or disadvantageous. Instead, we require full knowledge of the boundary layer profile $\hat{\delta}_{0.99}(\hat{x})$, inserted into (\ref{CwG2}).

We have also tried fitting the shape of the boundary layer $\hat{\delta}_{0.99}(\hat{x})$ to match together the theoretical wave drag coefficient (\ref{CwG2}) and the experimental data in figure \ref{exp}(a), using a numerical least-squares optimisation method. We keep the details of this optimisation in Appendix \ref{appC}, but we display the results in figure \ref{anal}(d). The corresponding boundary layer thicknesses are displayed in (a) with blue dotted curves, and the mean thickness in (b). 
The numerical optimisation matches the theoretical data with the experimental data extremely well. The average relative error is approximately $5\%$, compared to $18\%$ using the $k$-$\omega$ SST boundary layer in (c), and $26\%$ using no boundary layer at all.

The numerical optimisation finds a slightly larger boundary layer thickness than extracted from the $k$-$\omega$ SST model. There are several possible reasons for this discrepancy. For example, the $k$-$\omega$ SST model may under-predict the growth of the turbulent boundary layer. Or, perhaps the fitted boundary layer naturally corresponds to a larger thickness than the $99\%$ definition.
In any case, it is evident from the boundary layer profiles in figure \ref{anal}(a) that the fitted boundary layer and the $k$-$\omega$ SST boundary layer are similar in shape. This suggests that our modification to Michell's theory is appropriate, and provided good knowledge of the boundary layer profile, the wave drag on an asymmetric body can be predicted much more reliably than the original formulation (\ref{CwG}). 
However, we acknowledge that this is an effective approach, and does not manifest a complete description of the flow in the boundary layer. 
%Given its simplicity, this effective approach works surprisingly well,  

\section{Influence of the depth of motion}

\begin{figure}
\centering
\begin{tikzpicture}[scale=0.5]
\node at (0,0) {\includegraphics[width=0.45\textwidth]{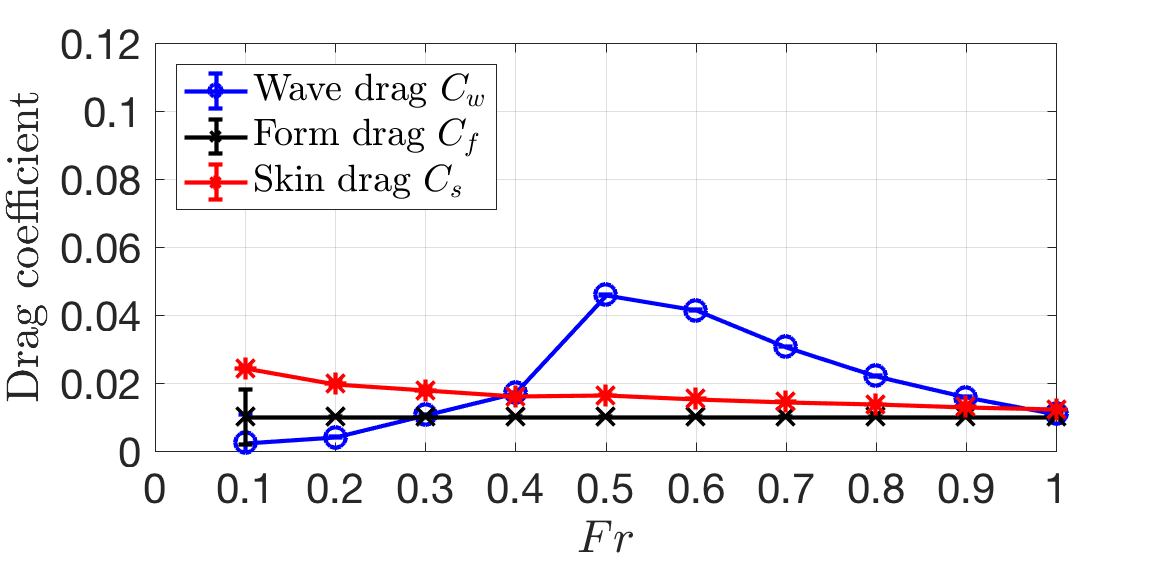}};
\node at (1.8,3.6){{\scalebox{-1}[1]{\includegraphics[width=0.15\textwidth]{figs/3d_hulls_hull_5_new}}}};
\begin{axis}[domain=0:50,hide axis, scale only axis, width=0.7\textwidth,
     height=0.01\textwidth, samples=100, at={(-0.4\textwidth,0.26\textwidth)}]
      \addplot[mark=none,color=blue,very thick]{0.5*sin(50*x)};
    \end{axis}
\node at (2,0.7) { \textit{Wave dominated}};
\node at (3,2.9) { $\epsilon<0$};
\draw[line width=1,->] (3.5,3.6) -- (4.5,3.6);
\draw[line width=1,dashed] (-0.8,-1.0) --(-0.8,0.2) -- (3.8,0.2)-- (3.8,-1.0);
\node at (-2.0,4.5) {\large \bf $\boldsymbol{d=0.5}$};
\node at (-5.0,4.2) {(a)};
\end{tikzpicture}
%\begin{overpic}[width=0.45\textwidth]{figs/both_drag_05_b}
%\put (-10,60) {(b)}
%\put (-15,50) {\large \bf $\boldsymbol{d=0.5}$}
%\put (45,32) {  \textit{Wave dominated}}
%\end{overpic}\\
\begin{tikzpicture}[scale=0.5]
\node at (0,0) {\includegraphics[width=0.45\textwidth]{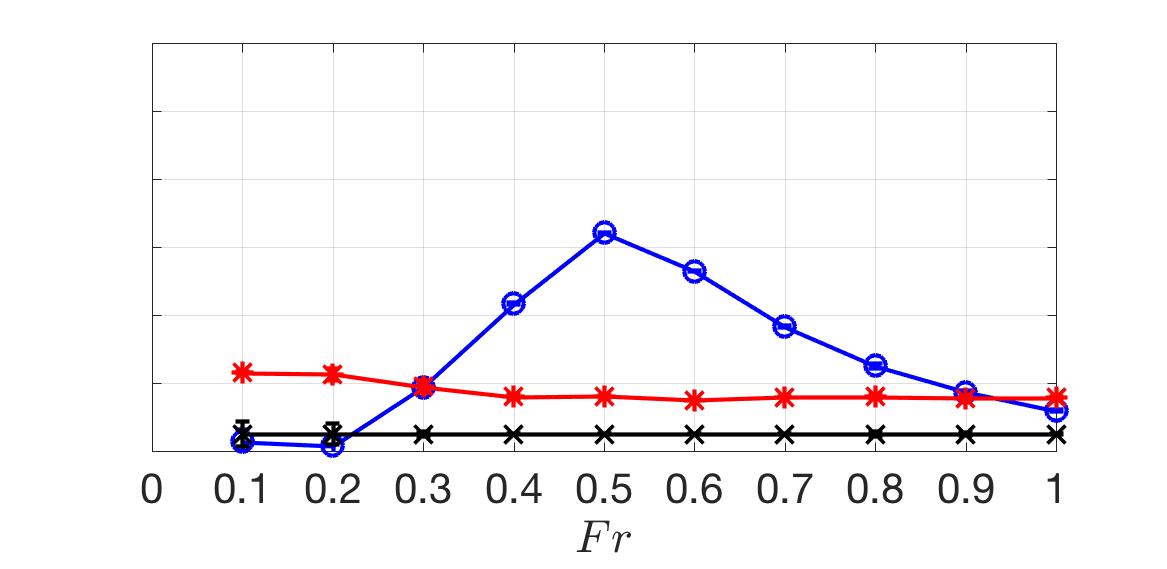}};
\node at (1.8,3.6){\includegraphics[width=0.15\textwidth]{figs/3d_hulls_hull_5_new}};
\begin{axis}[domain=0:50,hide axis, scale only axis, width=0.7\textwidth,
     height=0.02\textwidth, samples=100, at={(-0.4\textwidth,0.26\textwidth)}]
      \addplot[mark=none,color=blue,very thick]{0.5*sin(50*x)};
    \end{axis}
\node at (2,1.3) { \textit{Wave dominated}};
\node at (3,2.9) { $\epsilon>0$};
\draw[line width=1,->] (3.5,3.6) -- (4.5,3.6);
\draw[line width=1,dashed] (-1.6,-1.0) --(-1.6,0.8) -- (3.8,0.8)-- (3.8,-1.0);
\node at (-5.0,4.2) {(b)};
\end{tikzpicture}\\
\vspace{0.2cm}
%\begin{overpic}[width=0.45\textwidth]{figs/both_drag_2_f}
%\put (32,22) {  \textit{Skin/Form dominated}}
%\put (-10,60) {(a)}
%\put (40,65) {\bf $\boldsymbol{d=0.5}$}
%\end{overpic}
\begin{tikzpicture}[scale=0.5]
\node at (0,0) {\includegraphics[width=0.45\textwidth]{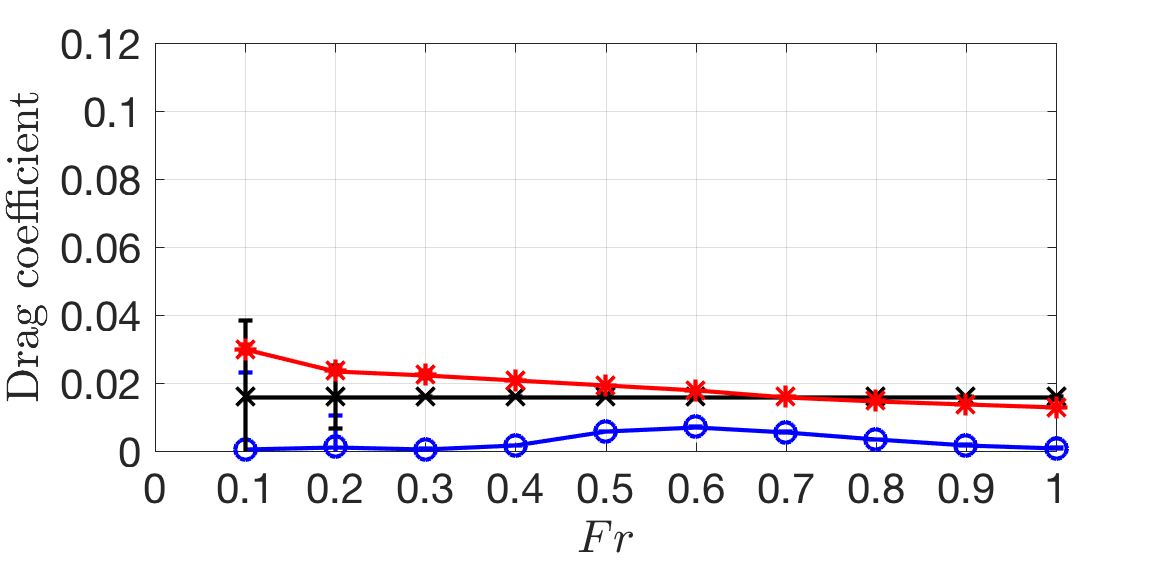}};
\node at (1.8,3.6){{\scalebox{-1}[1]{\includegraphics[width=0.15\textwidth]{figs/3d_hulls_hull_5_new}}}};
\begin{axis}[domain=0:50,hide axis, scale only axis, width=0.7\textwidth,
     height=0.01\textwidth, samples=100, at={(-0.4\textwidth,0.38\textwidth)}]
      \addplot[mark=none,color=blue,very thick]{0.5*sin(50*x)};
    \end{axis}
\node at (1.6,-0.1) { \textit{Skin/Form dominated}};
\node at (3,2.9) { $\epsilon<0$};
\draw[line width=1,->] (3.5,3.6) -- (4.5,3.6);
\node at (-2.0,4.0) {\large \bf $\boldsymbol{d=2.0}$};
\node at (-5.0,4.2) {(c)};
\end{tikzpicture}
%\begin{overpic}[width=0.45\textwidth]{figs/both_drag_2_b}
%\put (36,22) {  \textit{Skin dominated}}
%\put (-10,60) {(b)}
%\put (-15,50) {\large \bf $\boldsymbol{d=2.0}$}
%\end{overpic}
\begin{tikzpicture}[scale=0.5]
\node at (0,0) {\includegraphics[width=0.45\textwidth]{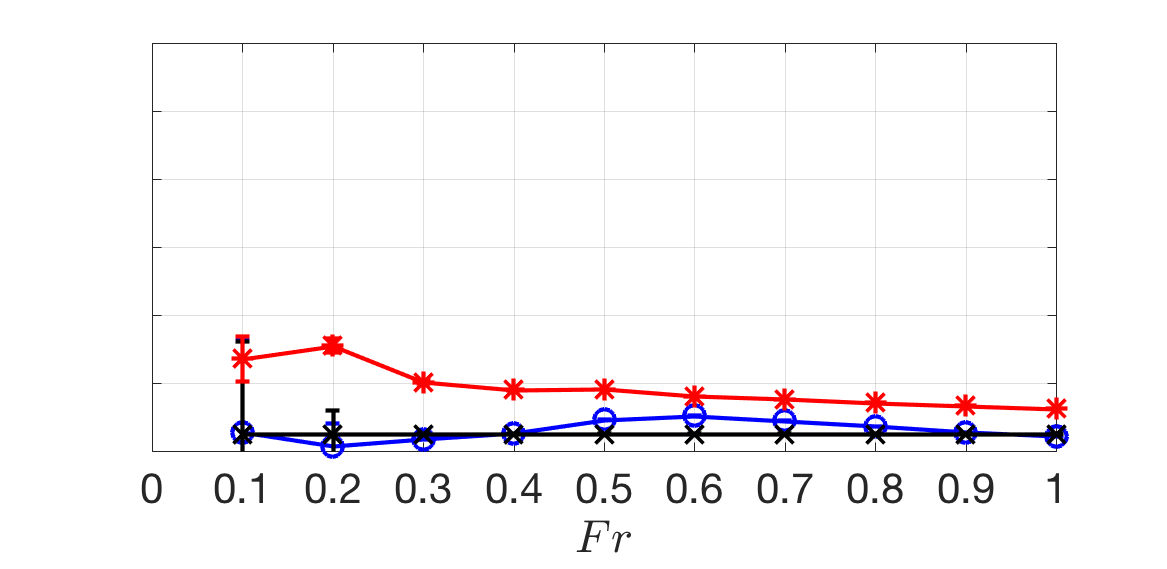}};
\node at (1.8,3.6){{{\includegraphics[width=0.15\textwidth]{figs/3d_hulls_hull_5_new}}}};
\begin{axis}[domain=0:50,hide axis, scale only axis, width=0.7\textwidth,
     height=0.01\textwidth, samples=100, at={(-0.4\textwidth,0.38\textwidth)}]
      \addplot[mark=none,color=blue,very thick]{0.5*sin(50*x)};
    \end{axis}
\node at (2,-0.1) { \textit{Skin dominated}};
\node at (3,2.9) { $\epsilon>0$};
\draw[line width=1,->] (3.5,3.6) -- (4.5,3.6);
\node at (-5.0,4.2) {(d)};
\end{tikzpicture}
\caption{Results from the $k$-$\omega$ SST model, where the numerically computed drag coefficient is decomposed into wave, form and skin components $C_w,\,C_f,\,C_s$. (a, b) Hull 5 from the slender family at depth $d=0.5$ with $\epsilon<0$ (a) and $\epsilon>0$ (b). (c, d) Hull 5 from the slender family at depth $d=2.0$ with $\epsilon<0$ (c) and $\epsilon>0$ (d).  \label{numdisc} }
\end{figure}

In figure \ref{2depths} we showed that asymmetry can be advantageous or disadvantageous, depending on whether the body is near or far away from the air-water interface. This is explained by the relative importance of wave and form drag. For large depths, form drag dominates over wave drag, such that positive asymmetry is favourable. On the other hand, at shallower depths wave drag dominates over form drag, such that negative asymmetry is better. 
Whilst we have shown that, using a wind tunnel, it is possible to measure wave drag independently from the combined total of skin and form drag, it is quite challenging to treat form and skin components independently, at least from an experimental point of view. 
However, with numerical simulations, such as the $k$-$\omega$ SST model, this is relatively straightforward. 
Being able to decompose the drag into these three components is very useful when comparing their relative magnitudes.

From the results of our $k$-$\omega$ SST model, it is possible to extract the time-averaged pressure and viscous stresses integrated over the hull surface. Since form and wave drag result from a pressure force, these are lumped together to form a pressure coefficient, which we denote $C_p$. This is given in dimensional terms as
\beq
C_{p}:=C_f+C_w =\frac{1}{\rho U^2 \Omega^{2/3}} \int_S  \lb  p\, \boldsymbol{I} \cdot \hat{\boldsymbol{n}} \rb \cdot \hat{\boldsymbol{\i} }\,\mathrm{d} S ,\label{numsplit}
\eeq
where $\hat{\boldsymbol{\i} }$ is the unit vector in the $x$ direction, $\hat{\boldsymbol{n}}$ is the unit outward-pointing normal to the hull surface $S$, and $\boldsymbol{I}$ is the identity matrix.
Similarly, the skin drag can be calculated from the viscous stress component
\beq
C_s =\frac{1}{\rho U^2 \Omega^{2/3}} \int_S  \mu  \lb  \lb \nabla \boldsymbol{u}+\nabla \boldsymbol{u}^T \rb  \cdot \hat{\boldsymbol{n}} \rb \cdot \hat{\boldsymbol{\i} }\,\mathrm{d} S.
\eeq
%In fact, the close to the hull wall, the viscosity is much larger than the eddy viscosity, 

It is not immediately obvious how to split $C_f$ and $C_w$ in (\ref{numsplit}). However, this can be achieved by noting two particular properties of $C_f$ and $C_w$. Firstly, in the limit $\mathrm{\it Fr}\rightarrow 0$ or $\mathrm{\it Fr}\rightarrow\infty$, we expect $C_w\rightarrow 0$ \citep{michell1898xi,tuck1989wave}. Secondly, we do not expect $C_f$ to depend strongly on the Reynolds number, and hence the Froude number (see figure \ref{exp}(b)). Hence, the form drag can be extracted as
\beq
C_f=\lim_{\mathrm{\it Fr}\rightarrow 0}C_p(\mathrm{\it Fr}),\label{wavelim}
\eeq
and, consequently, the wave drag can be approximated as
\beq
C_w(\mathrm{\it Fr})=C_p(\mathrm{\it Fr})-C_f.
\eeq
We are unable to compute $C_d$ in the limit $\mathrm{\it Fr}\rightarrow 0$ since the $k$-$\omega$ SST model is only valid for turbulent flows. Hence, we approximate (\ref{wavelim}) by averaging $C_d$ over a few small values of $\mathrm{\it Fr}$. We think this is an acceptable approach, since from figure \ref{exp}(c) we can see that Michell's theory predicts rapid decay of $C_w$ for $\mathrm{\it Fr}<0.2$.

%\todonew{Not sure exactly how true this statement is}

\begin{figure}
\centering
\vspace{0.5cm}
\begin{overpic}[width=0.45\textwidth]{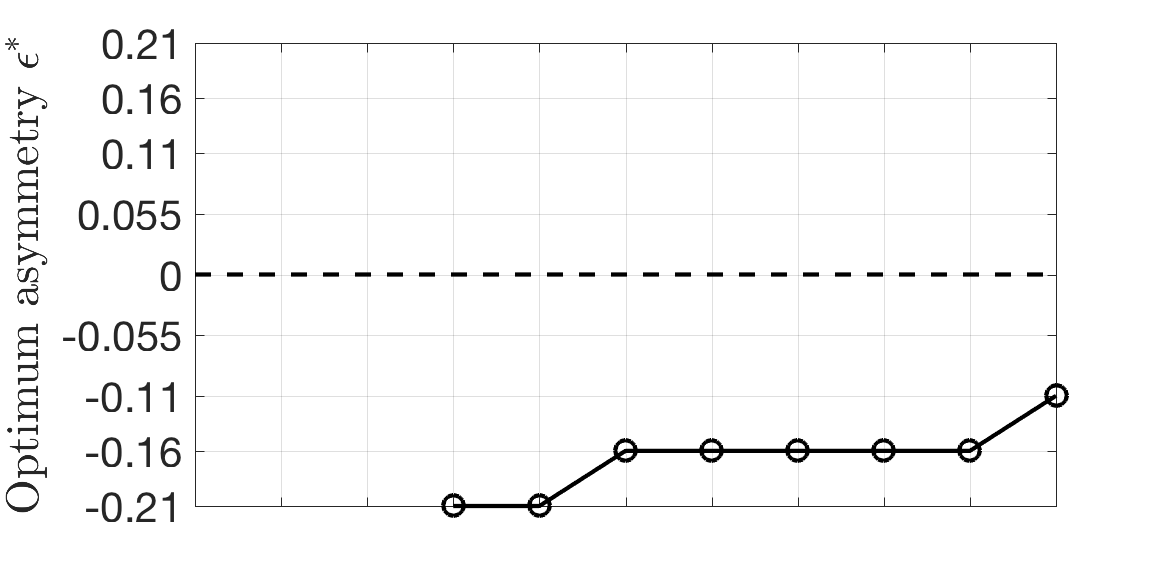}
\put (-5,50) {(a)}
\put (45,50) {$\boldsymbol{d=0.5}$}
%\put (90,27) {\includegraphics[width=0.1\textwidth]{figs/snap}}
%\put (45,12) {\includegraphics[width=0.1\textwidth]{figs/snap2}}
\end{overpic}
\begin{overpic}[width=0.45\textwidth]{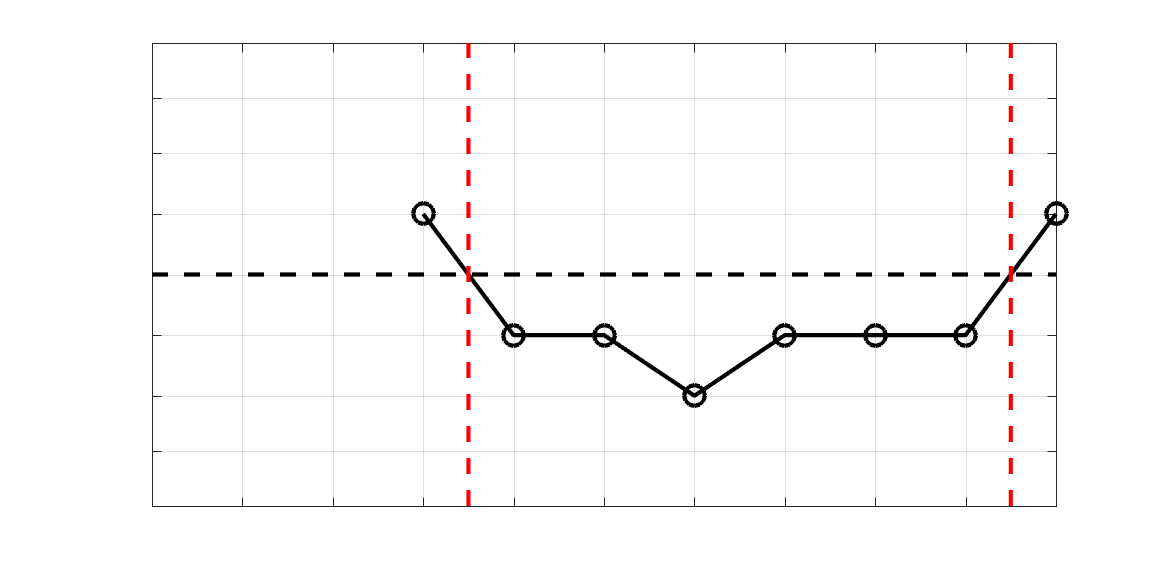}
\put (-10,43) {\includegraphics[width=0.075\textwidth]{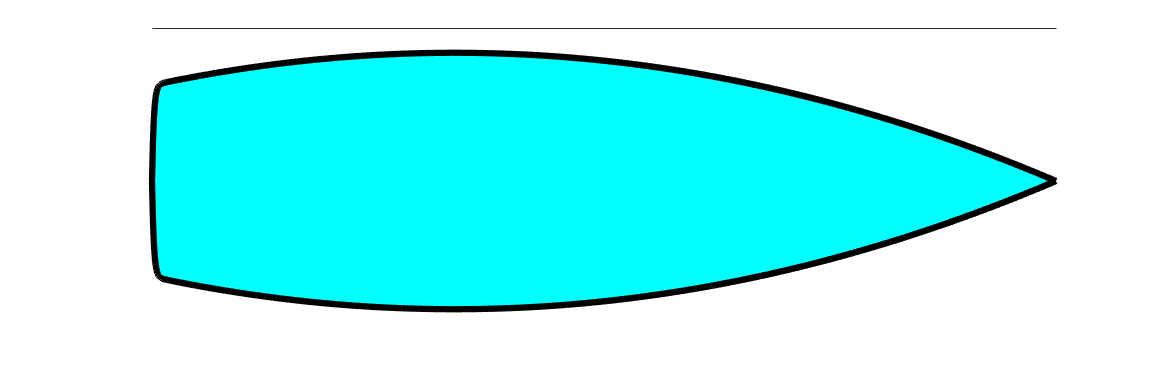}}
\put(-5,46){\color{black}\vector(1,0){17}}
\put (-10,38) {\includegraphics[width=0.075\textwidth]{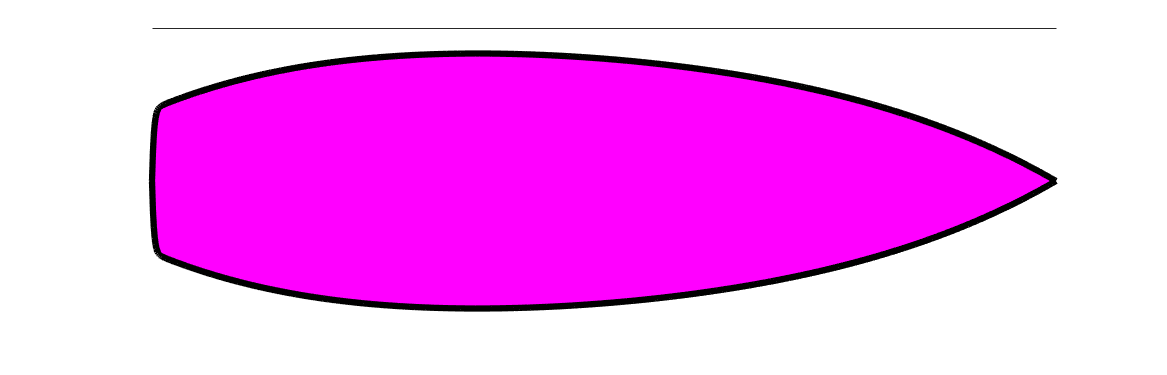}}
\put(-5,41){\color{black}\vector(1,0){17}}
\put (-10,33) {\includegraphics[width=0.075\textwidth]{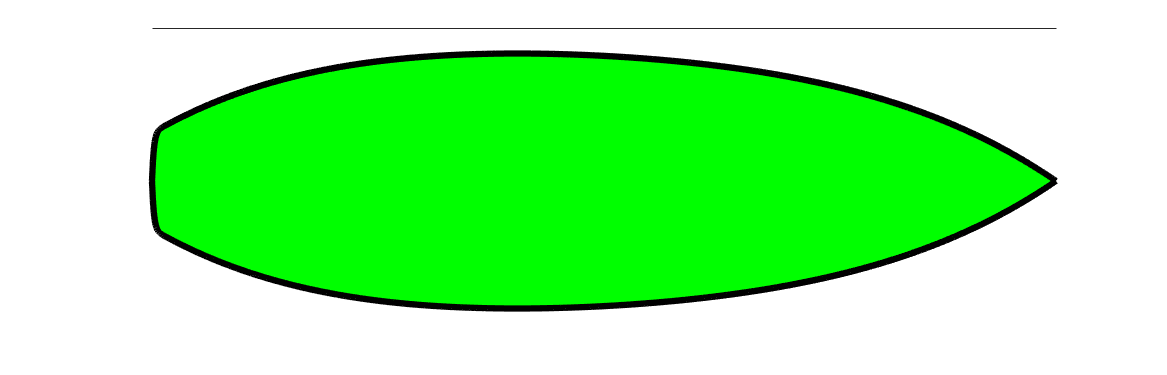}}
\put(-5,36){\color{black}\vector(1,0){17}}
\put (-10,28) {\includegraphics[width=0.075\textwidth]{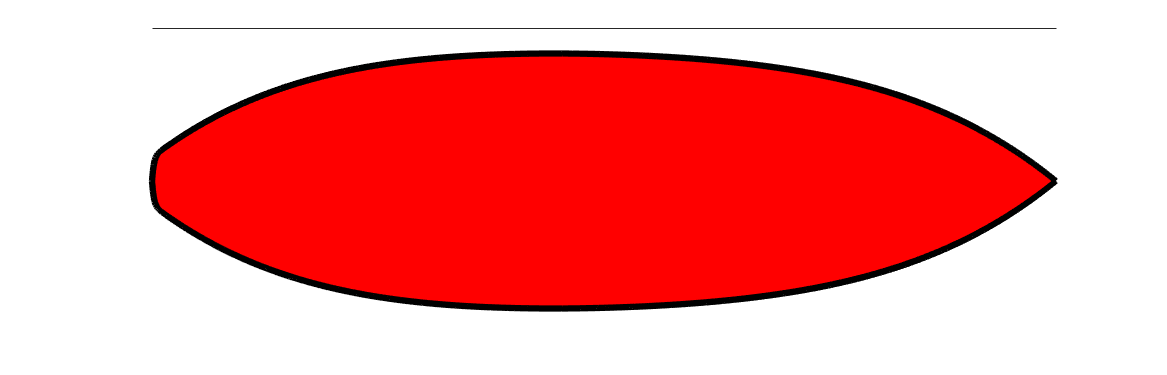}}
\put(-5,31){\color{black}\vector(1,0){17}}
\put (-10,23) {\includegraphics[width=0.075\textwidth]{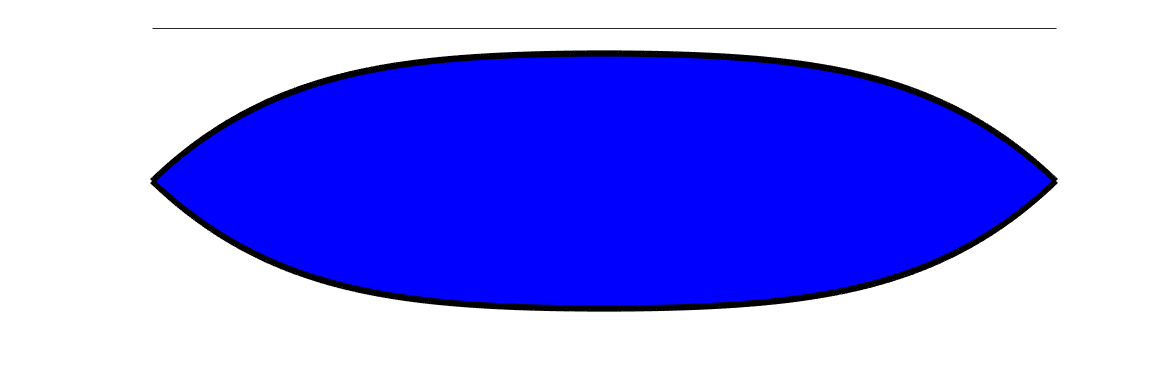}}
\put(-5,26){\color{black}\vector(1,0){17}}
\put (-10,18) {\scalebox{-1}[1]{\includegraphics[width=0.075\textwidth]{figs/new_hulls_hull_2_new}}}
\put(-5,21){\color{black}\vector(1,0){17}}
\put (-10,13) {\scalebox{-1}[1]{\includegraphics[width=0.075\textwidth]{figs/new_hulls_hull_3_new}}}
\put(-5,16){\color{black}\vector(1,0){17}}
\put (-10,8) {\scalebox{-1}[1]{\includegraphics[width=0.075\textwidth]{figs/new_hulls_hull_4_new}}}
\put(-5,11){\color{black}\vector(1,0){17}}
\put (-10,3) {\scalebox{-1}[1]{\includegraphics[width=0.075\textwidth]{figs/new_hulls_hull_5_new}}}
\put(-5,6){\color{black}\vector(1,0){17}}
\put (5,50) {(b)}
\put (45,50) {$\boldsymbol{d=1.0}$}
\end{overpic}\\
\vspace{0.3cm}
\begin{overpic}[width=0.45\textwidth]{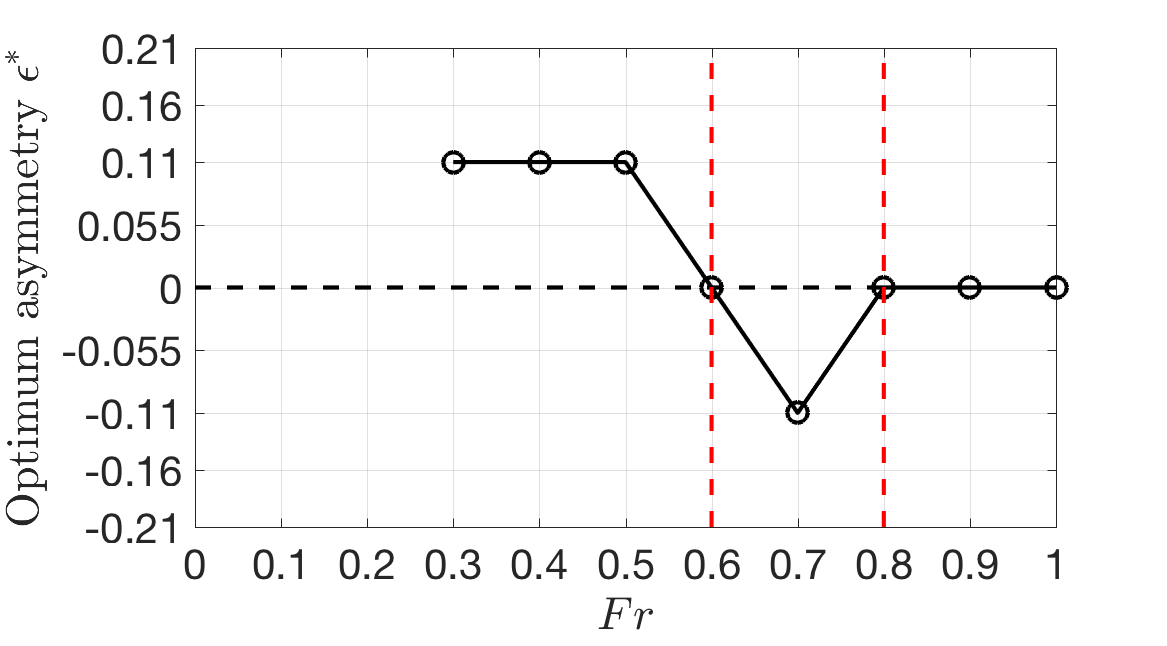}
\put (-5,54) {(c)}
\put (45,54) {$\boldsymbol{d=1.25}$}
\end{overpic}
\begin{overpic}[width=0.45\textwidth]{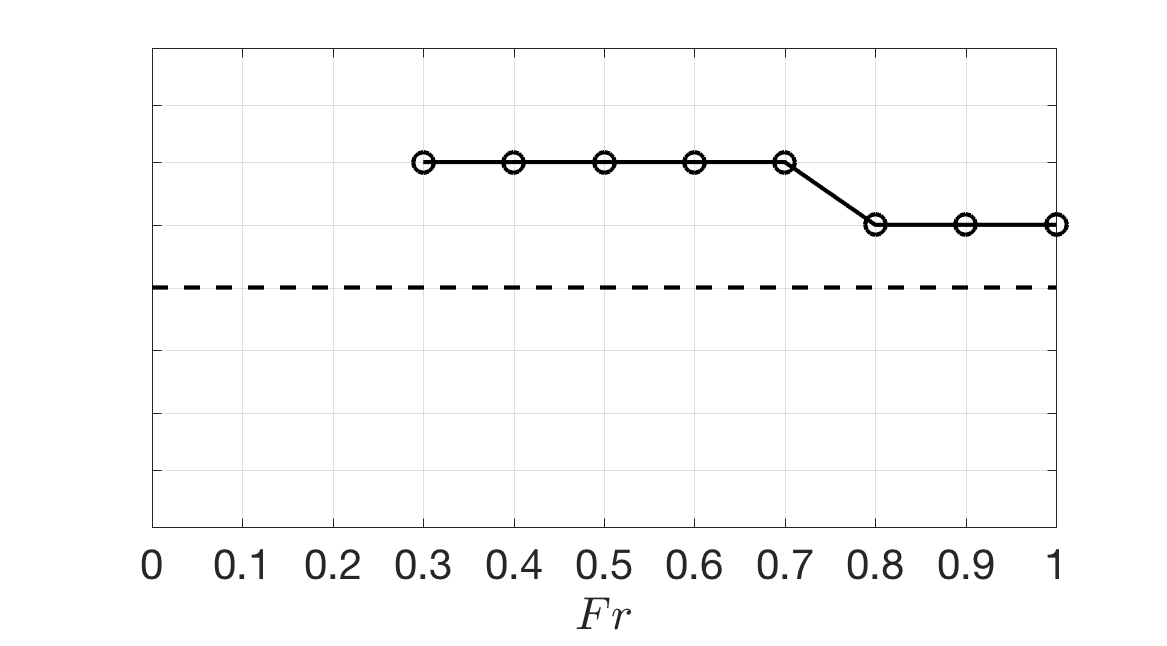}
\put (5,54) {(d)}
\put (45,54) {$\boldsymbol{d=2.0}$}
\put (-10,47) {\includegraphics[width=0.075\textwidth]{figs/new_hulls_hull_5_new}}
\put(-5,50){\color{black}\vector(1,0){17}}
\put (-10,42) {\includegraphics[width=0.075\textwidth]{figs/new_hulls_hull_4_new}}
\put(-5,45){\color{black}\vector(1,0){17}}
\put (-10,37) {\includegraphics[width=0.075\textwidth]{figs/new_hulls_hull_3_new}}
\put(-5,40){\color{black}\vector(1,0){17}}
\put (-10,32) {\includegraphics[width=0.075\textwidth]{figs/new_hulls_hull_2_new}}
\put(-5,35){\color{black}\vector(1,0){17}}
\put (-10,27) {\includegraphics[width=0.075\textwidth]{figs/new_hulls_hull_1_new}}
\put(-5,30){\color{black}\vector(1,0){17}}
\put (-10,22) {\scalebox{-1}[1]{\includegraphics[width=0.075\textwidth]{figs/new_hulls_hull_2_new}}}
\put(-5,25){\color{black}\vector(1,0){17}}
\put (-10,17) {\scalebox{-1}[1]{\includegraphics[width=0.075\textwidth]{figs/new_hulls_hull_3_new}}}
\put(-5,20){\color{black}\vector(1,0){17}}
\put (-10,12) {\scalebox{-1}[1]{\includegraphics[width=0.075\textwidth]{figs/new_hulls_hull_4_new}}}
\put(-5,15){\color{black}\vector(1,0){17}}
\put (-10,7) {\scalebox{-1}[1]{\includegraphics[width=0.075\textwidth]{figs/new_hulls_hull_5_new}}}
\put(-5,10){\color{black}\vector(1,0){17}}
\end{overpic}
%\begin{overpic}[width=0.4\textwidth]{figs/Fr_crit.png}
%\put (82,18) {\color{red} \rotatebox{90}{$\boldsymbol{\epsilon^*>0}$}}
%\put (20,18) {\color{red} \rotatebox{90}{$\boldsymbol{\epsilon^*<0}$}}
%\put (-10,35) {(e)}
%\end{overpic}
\caption{(a)-(d) Optimal asymmetry $\epsilon^*$ as a function of Froude number for depths $d=0.5,1.0,1.25,2.0$, using experimental measurements of the bluff family of shapes. \label{bluffexp}}
\end{figure}

Using the above method, in figure \ref{numdisc} we plot the wave, skin and form drag coefficients for hull $5$ from the slender family with both positive and negative asymmetry for $\mathrm{\it Fr}\in[0.1,1.0]$ and two different depths $d=0.5,\,2.0$. When the hull moves close to the interface ($d=0.5$) the wave drag coefficient is larger than the skin and form drag coefficients for Froude numbers in the approximate range $\mathrm{\it Fr}\in[0.4,0.9]$, for both $\epsilon>0$ and $\epsilon<0$. For Froude numbers outside that range the wave drag coefficient decays and is comparable to the other drag components. Hence, for $\mathrm{\it Fr}\in[0.4,0.9]$ we expect a negative $\epsilon$ to be advantageous, whereas for large or small $\mathrm{\it Fr}$, we expect a positive $\epsilon$ to be advantageous.

When the body moves at the larger depth $d=2.0$ we can see that the wave drag component is never the largest component, regardless of Froude number. For $\epsilon<0$ form and skin drag are both of the same order of magnitude, but considerably larger than wave drag. This is because, for a body with a bluff trailing edge, boundary layer separation results in significant form drag. However, for $\epsilon>0$, since the body is more streamlined, form drag is smaller than skin drag. Hence, in this case, the form drag coefficient is smaller than the skin drag coefficient for all Froude numbers. Therefore, at large depths positive $\epsilon$ is advantageous.

The most obvious next question is the following: for a given depth and Froude number, which asymmetry is optimal? To answer this question we turn our attention to the bluff family of shapes. This family is more suitable than the slender family since there is a greater difference in drag coefficient between each hull shape within the family, giving us more granularity.

\begin{comment}
\begin{figure}
\centering
\begin{overpic}[width=0.45\textwidth]{figs/cylinder_bl}
\end{overpic}
\begin{overpic}[width=0.45\textwidth]{figs/cylinder_drag}
\end{overpic}
\caption{Boundary layer on a cylinder}
\end{figure}
\end{comment}

%\clearpage

%\clearpage

In figure \ref{bluffexp} we plot the optimum asymmetry, which we denote $\epsilon^*$, as a function of Froude number and depth, as measured in our experiments. Using $5$ hull shapes from the bluff family, for each of $\epsilon>0$ and $\epsilon<0$, there are a total of 9 possible values of the asymmetry parameter in the range $\epsilon\in [-0.21,0.21]$.
We see that for $d=0.5$ the optimum asymmetry is negative for all Froude numbers, whereas for $d=2.0$ the optimum asymmetry is positive. However for intermediate depths $d=1.0$ and $d=1.25$, the optimum asymmetry is sometimes positive and sometimes negative, depending on the Froude number range. In the case of $d=1.0$, $\epsilon<0$ is optimum for $\mathrm{\it Fr}\in[0.35,0.95]$, and $\epsilon>0$ is optimum for other $\mathrm{\it Fr}$ numbers. For $d=1.25$ there is a similar pattern, but the range of Froude numbers is smaller $\mathrm{\it Fr}\in[0.6,0.8]$. These results are qualitatively consistent with those in figure \ref{numdisc}, where $C_w$ dominates for a range of Froude numbers, but only at smaller depths (Note, however, that we use a different family of shapes between figures \ref{numdisc} and \ref{bluffexp}).

%\todonew{Currently Figure \ref{bluffexp} is quantitatively inconsistent with Figure \ref{numdisc} because at $d=0.5$ we should see a switch for small and large Fr numbers.}

% CONCLUSIONS

\section{Conclusions}

We have addressed the effect of front-back asymmetry on wave, form and skin drag for bodies moving at or near an interface. We have proposed two sets of body shapes, parameterised by a single quantity $\epsilon$, which measures the degree of the body asymmetry, and whose sign indicates whether the object has its pointed end at the leading or trailing edge. Using a combination of experimental, numerical and analytical approaches, we have illustrated how asymmetry can be advantageous or disadvantageous, depending on the submerged depth of the body and the Froude number. We have also proposed a simple modification of Michell's theory which enables the prediction of asymmetry effects using the turbulent boundary layer profile. The boundary layer is symmetry-breaking since, for an asymmetric body, it grows differently depending on the direction of motion.

% FUTURE WORK

For future work, the effect of top-bottom asymmetry could also be studied. Furthermore, we could use PIV to measure the boundary layer thickness experimentally. This would potentially enable more accurate measurement of the turbulent boundary layer thickness than
the $k$-$\omega$ SST model.
As a further step, a formal shape optimisation could be performed, where instead of considering shapes that are defined by a single asymmetry parameter, we would consider all continuous smooth shapes $\hat{f}(\hat{x})\in C^\infty [-1/2,1/2]$. 
The optimisation would require a model which has good capabilities in predicting each of the three drag coefficients, $C_w$, $C_f$ and $C_s$, especially when it comes to asymmetry effects. 
Here, we have presented a simple modification to Michell's theory for the wave drag $C_w$ that captures such effects, given knowledge of the boundary layer thickness. 
This simple modification is computationally inexpensive, which would make it ideal for optimisation.
Therefore, a reliable and computationally inexpensive model for the boundary layer (and hence $C_f$, $C_s$) would complement our modification, and allow for such an optimisation of the hull shape.

\begin{acknowledgements}
We thank Varvara Zhukovskaya and Bastien Garitaine for their contributions to the experiments conducted in this study. We also thank Renan Cuzon for useful discussions. We acknowledge the support from Ecole Polytechnique for the research program Sciences 2024.
\end{acknowledgements}

\appendix

\section{Mathematical expressions for the hull shapes}
\label{appA}

For the sake of reproducibility, in this section we give the expressions for the functions $\hat{f}(\hat{x})$ that we used for the slender and bluff families of shapes throughout the main text. In non-dimensional form, the slender family of shapes are given by
\beq
\hat{f}(\hat{x})=c_1\log \lb \frac{1+c_2}{e^{c_3 (\hat{x}-1/2)}+be^{-c_3 c_4 (\hat{x}-1/2)}} \rb,\label{slender_fun}
\eeq
where the coefficients $c_1,c_2,c_3,c_4$, for the $5$ different shapes are listed in table \ref{table1}. The bluff family of shapes are given by
\beq
\hat{f}(\hat{x})= c_1\lb c_3\lb 1/2 +\hat{x}\rb\lb 1 - e^{-c_4\lb 1/2 - \hat{x}\rb} \rb + \lb 1 - c_3\rb\lb 1/4 - \hat{x}^2\rb\lb \hat{x}^2 + c_2^2\rb\rb,\label{bluff_fun}
\eeq
and the corresponding coefficient values are listed in table \ref{table1}. The coefficients are chosen so that each of the shapes $\hat{f}(\hat{x})$ within the family have the following properties: $\hat{f}(\pm 1/2)=0$; $\max \{\hat{f}(\hat{x})\}=1/2$, and $\int_{-1/2}^{1/2} \hat{f}(\hat{x})\mathrm{d}\hat{x}=\hat{V}$, where the non-dimensional volume is $\hat{V}=0.31$ for the slender family and $\hat{V}=0.38$ for the bluff family.  The corresponding values of the asymmetry parameter $\epsilon$ are also listed in table \ref{table1}  for each shape.

\begin{table}
\centering
\begin{minipage}{0.45\textwidth}
\centering
Slender family\\
\begin{tabular}{|c|c|c|c|c|c|}
\hline
Shape & $c_1$ & $c_2$ & $c_3$ & $c_4$& $\epsilon$ \\
\hline
1 & 0.460& 0.030 & 3.500 & 1 & 0 \\
2 & 0.488 & 0.066 & 4.182 & 0.660 &0.057 \\
3 & 0.592 & 0.163 & 4.864 & 0.402 &0.113 \\
4 & 0.937 & 0.500 & 5.500 & 0.199 & 0.161\\
5 & 9.007 & 9.195 & 6.091 & 0.017 & 0.203\\
\hline
\end{tabular}
\end{minipage}
\begin{minipage}{0.45\textwidth}
\centering
Bluff family \\
\begin{tabular}{|c|c|c|c|c|c|}
\hline
Shape & $c_1$ & $c_2$ & $c_3$ & $c_4$ & $\epsilon$ \\
\hline
1 &  5.600 & 0.598  & 0          & 500 & 0 \\
2 &  4.060 & 0.674  & 0.023 & 500 & 0.053\\
3 &  2.810 & 0.778  & 0.067 & 500 & 0.108\\
4 &  1.953 & 0.901  & 0.144 & 500 &0.161 \\
5 & 0.376 & 54.972 & 0.999 & 500 & 0.215 \\
 \hline
\end{tabular}
\end{minipage}
\caption{List of the coefficients for the slender and bluff families of shapes (\ref{slender_fun}) and (\ref{bluff_fun}). The coefficients listed here correspond to $\epsilon>0$. The shapes with $\epsilon<0$ are achieved under the transformation $\hat{x}\rightarrow -\hat{x}$. \label{table1}}
\end{table}

\section{Additional drag calculations}
\label{appB}

In this section we display additional drag coefficients measured either using the tow-tank experiment, or the $k$-$\omega$ SST model, as described in the main text. 

\begin{figure}
\centering
\vspace{0.3cm}
\begin{overpic}[width=0.15\textwidth]{figs/all_logs_b}
\put  (0,100) {\textit{Colour scheme}}
\end{overpic}
\begin{overpic}[width=0.45\textwidth]{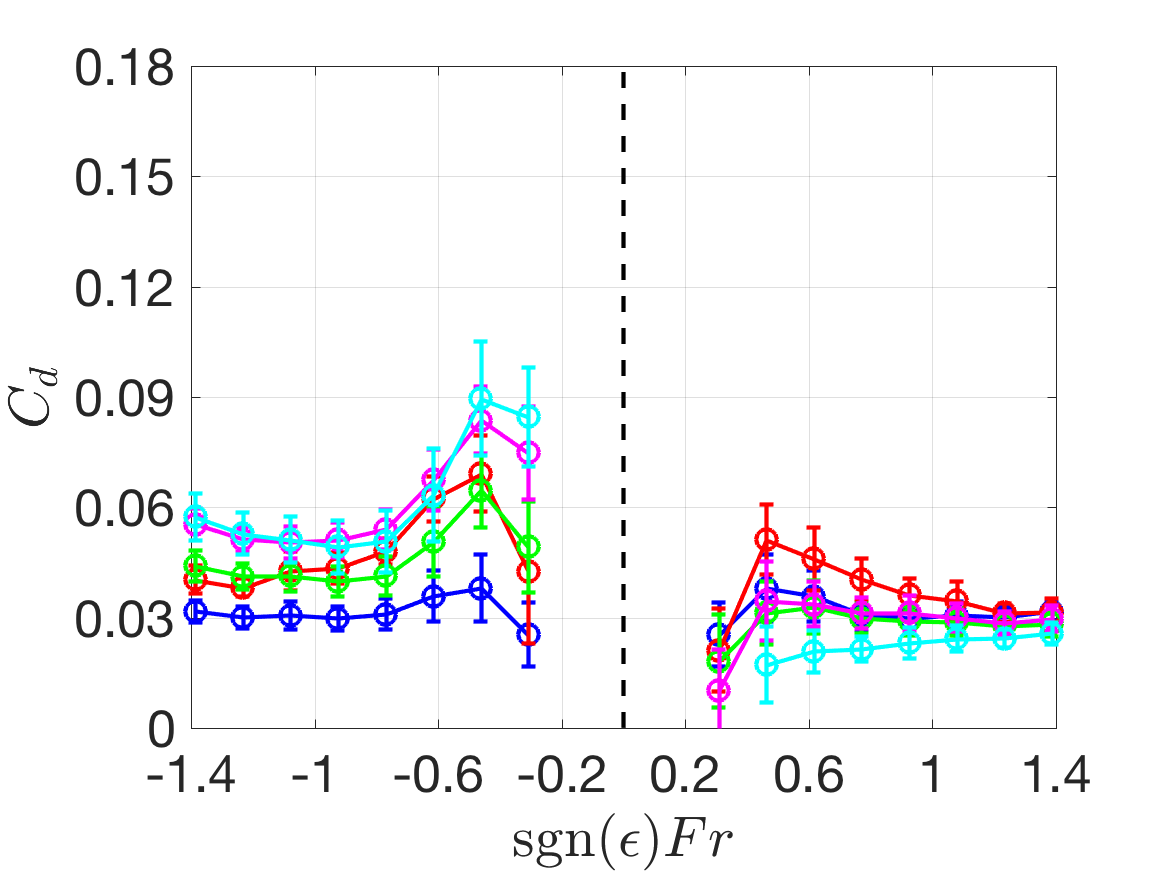}
\put (0,72) {(a)}
\put (42,72) {$\boldsymbol{d=0.25}$}
\put (15,80) {\large \bf Tow-tank measurements}
\end{overpic}
\begin{overpic}[width=0.15\textwidth]{figs/all_logs_f}
\put  (5,100) {\textit{Colour scheme}}
\end{overpic}\\
\vspace{0.3cm}
\begin{overpic}[width=0.45\textwidth]{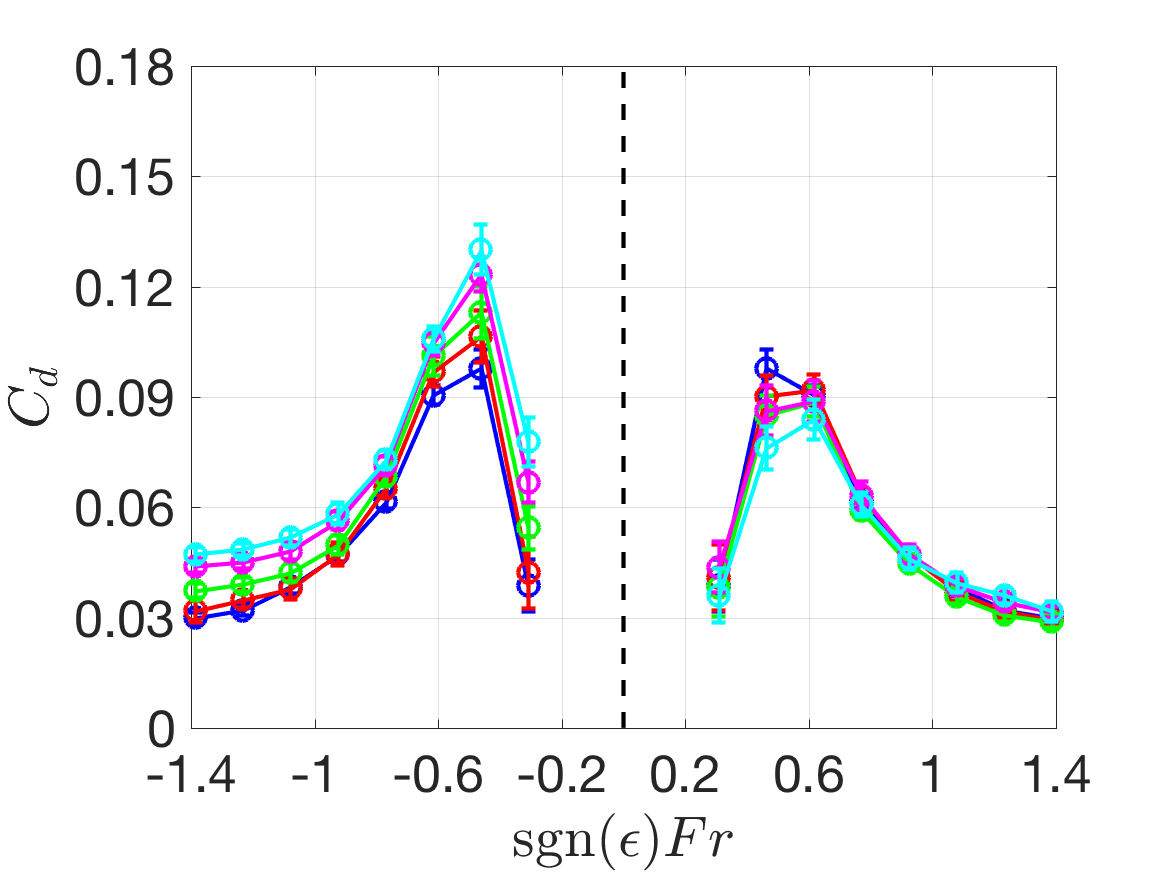}
\put (0,72) {(b)}
\put (42,72) {$\boldsymbol{d=0.75}$}
\end{overpic}
\begin{overpic}[width=0.45\textwidth]{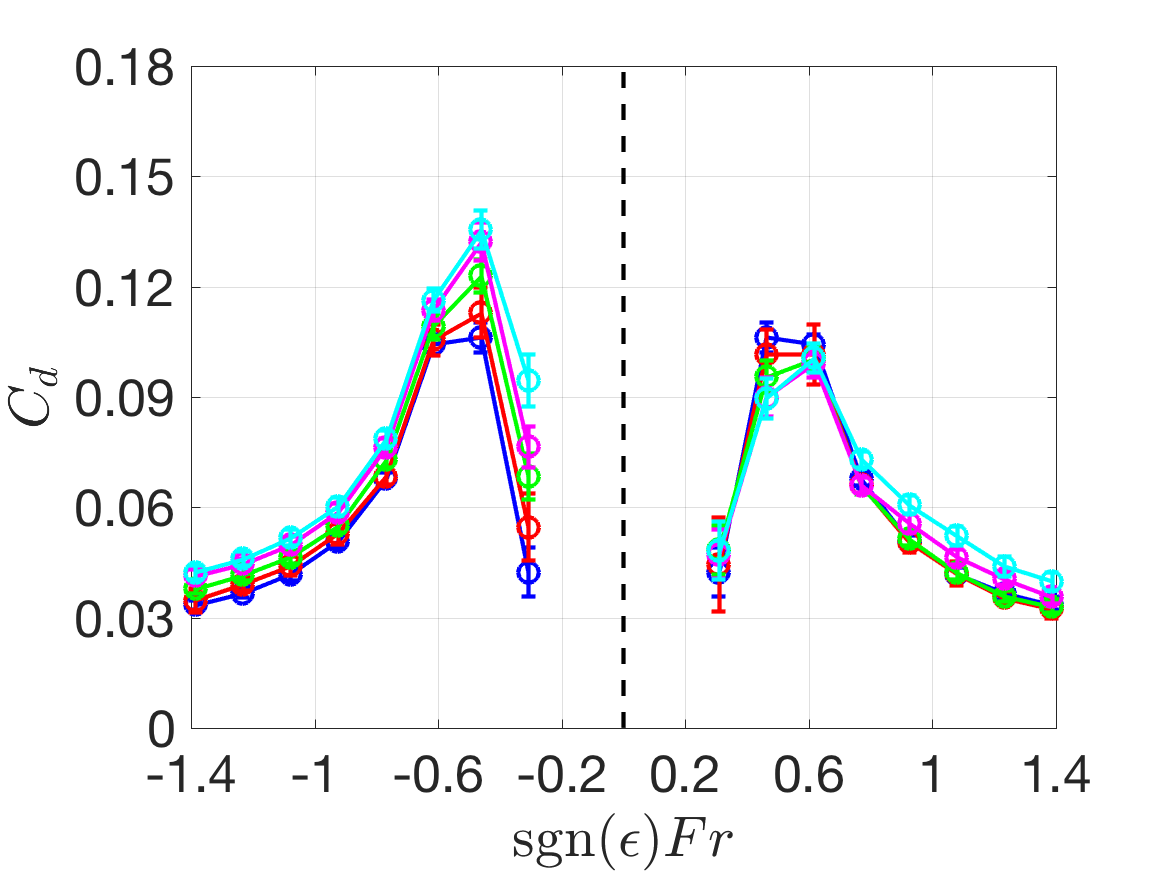}
\put (0,72) {(c)}
\put (42,72) {$\boldsymbol{d=1.0}$}
\end{overpic}
\caption{Additional tow-tank measurements of the drag coefficient $C_d$ for the slender family of shapes at various different depths. \label{app_slender}}
\end{figure}

\begin{figure}
\centering
\vspace{0.3cm}
\begin{overpic}[width=0.45\textwidth]{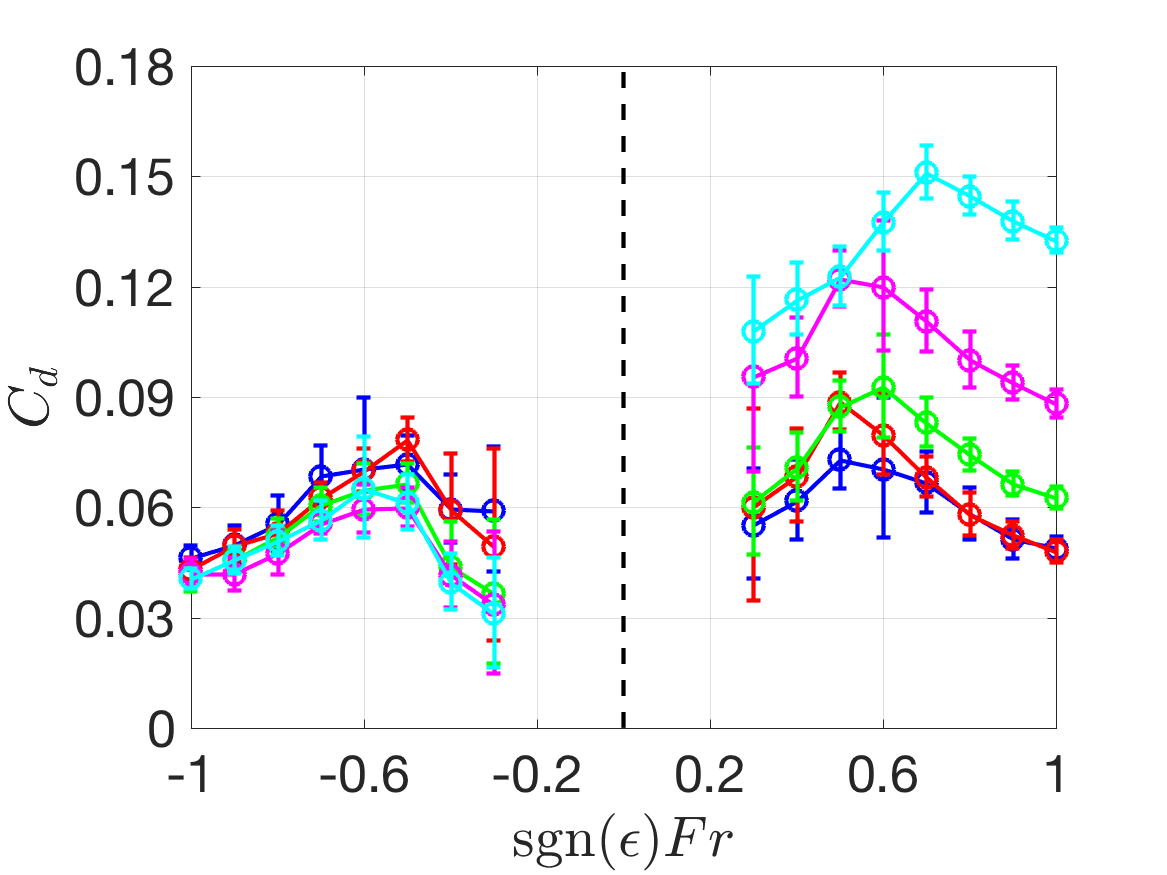}
\put (0,72) {(a)}
\put (42,72) {$\boldsymbol{d=0.5}$}
\end{overpic}
\begin{overpic}[width=0.45\textwidth]{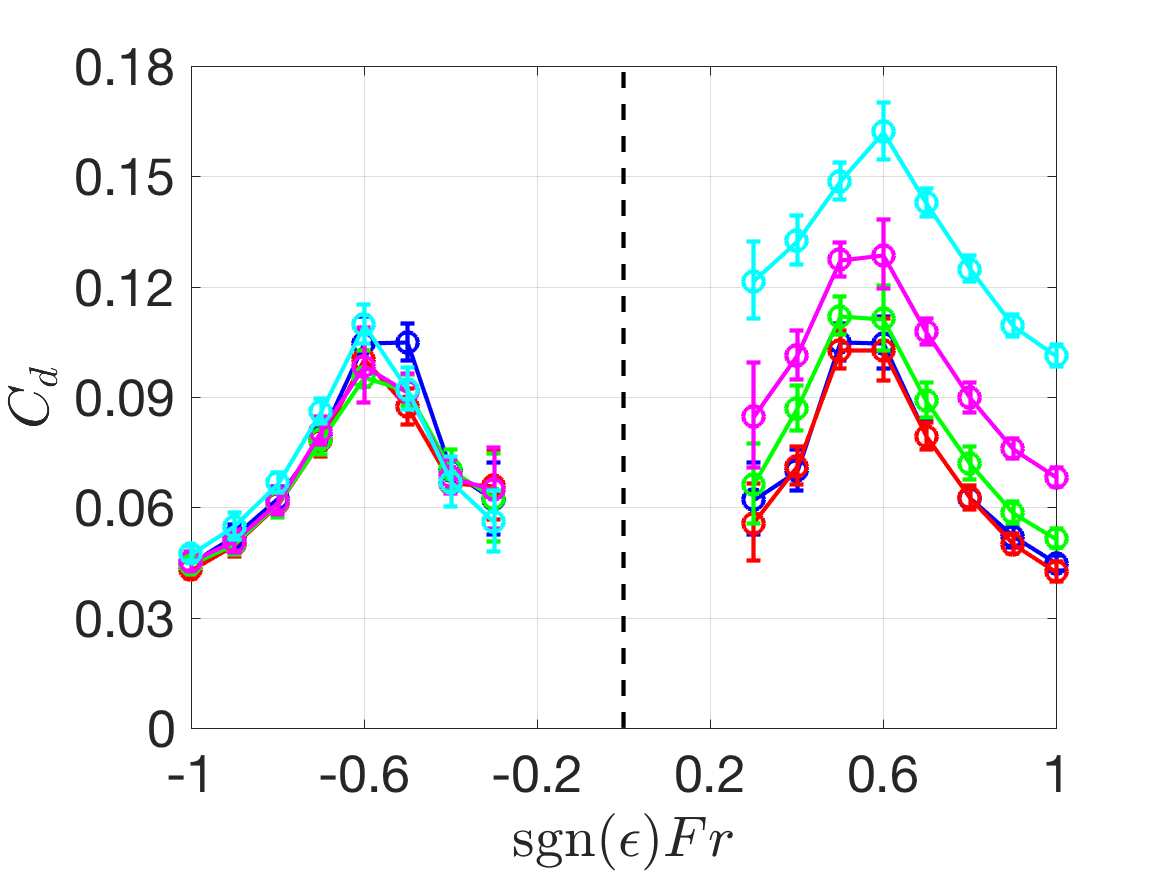}
\put (-42,80) {\large \bf Tow-tank measurements}
\put (0,72) {(b)}
\put (42,72) {$\boldsymbol{d=1.0}$}
\end{overpic}\\
\vspace{0.2cm}
\begin{overpic}[width=0.45\textwidth]{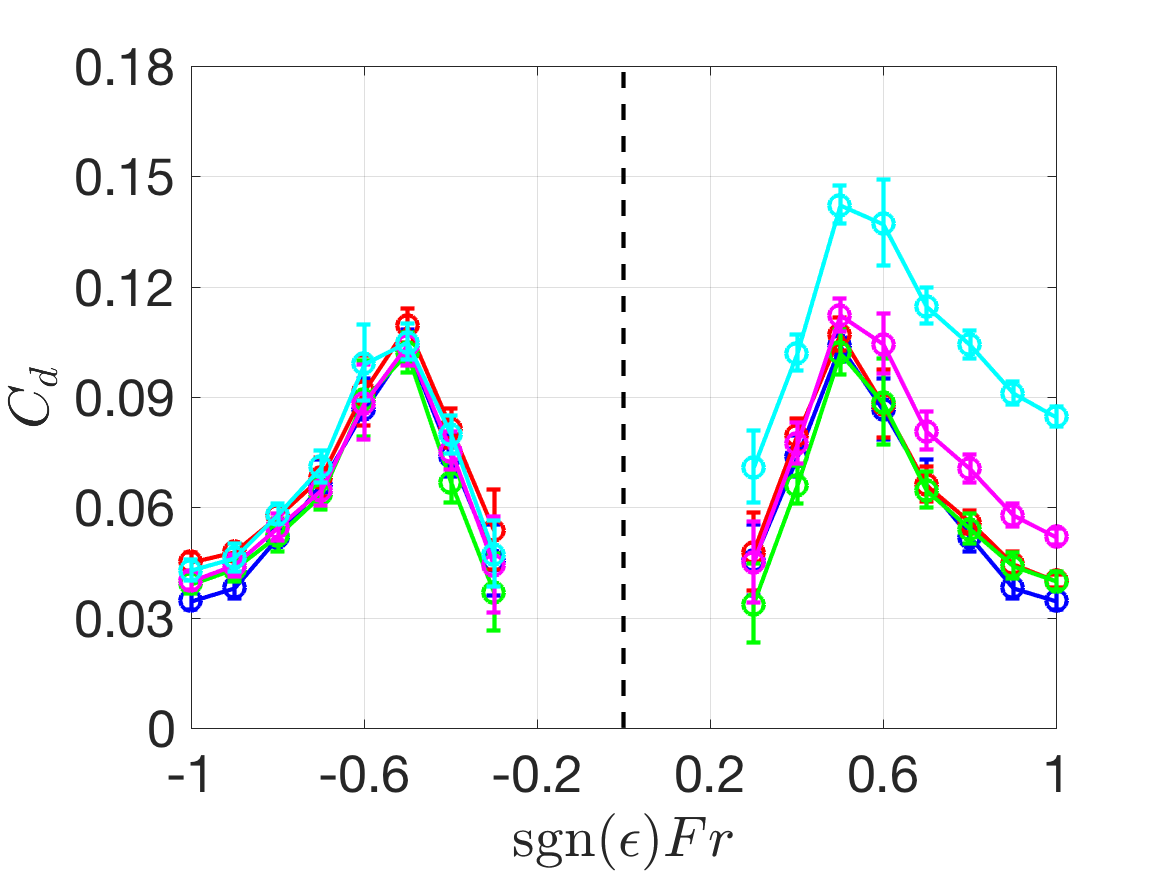}
\put (0,72) {(c)}
\put (42,72) {$\boldsymbol{d=1.25}$}
\end{overpic}
\begin{overpic}[width=0.45\textwidth]{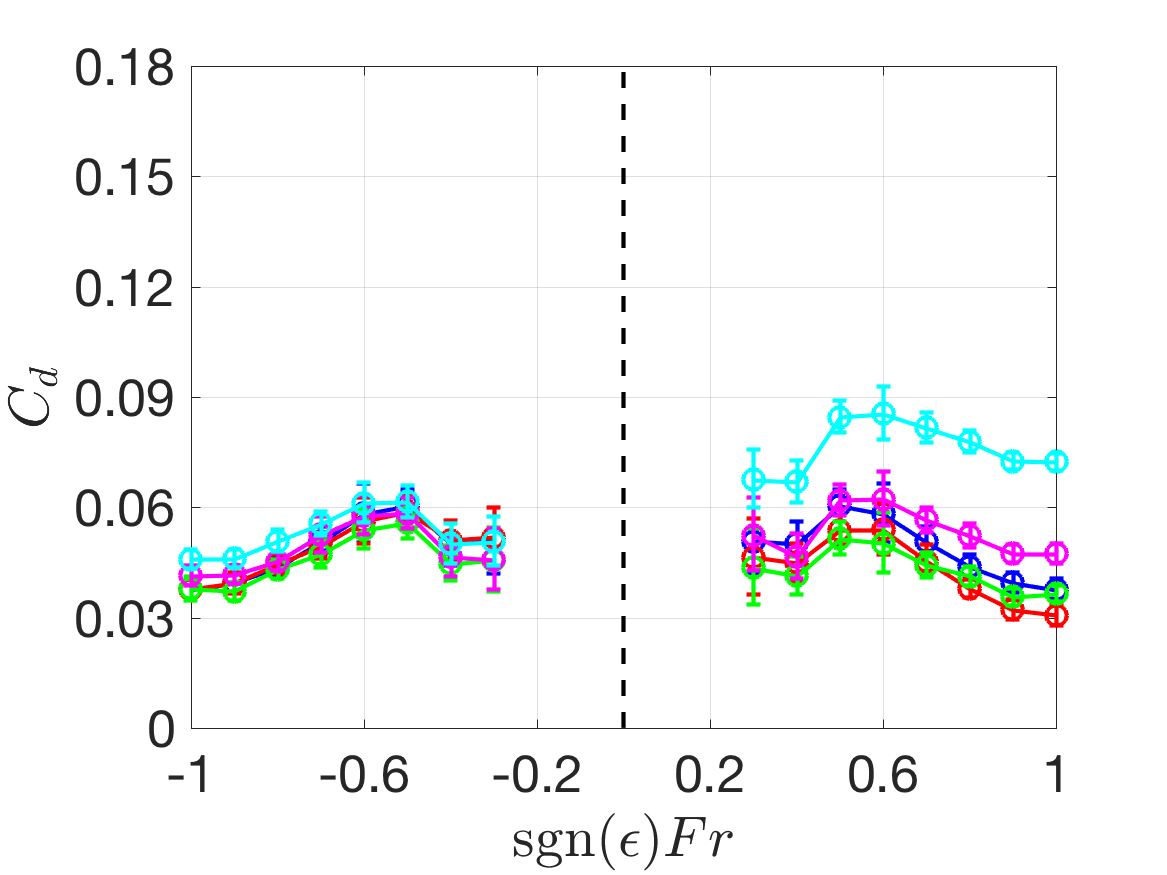}
\put (0,72) {(d)}
\put (42,72) {$\boldsymbol{d=2.0}$}
\end{overpic}\\
%\caption{Bluff family \label{app_bluff}}
%\end{figure}
%\begin{figure}
%\centering
\vspace{1cm}
\begin{overpic}[width=0.45\textwidth]{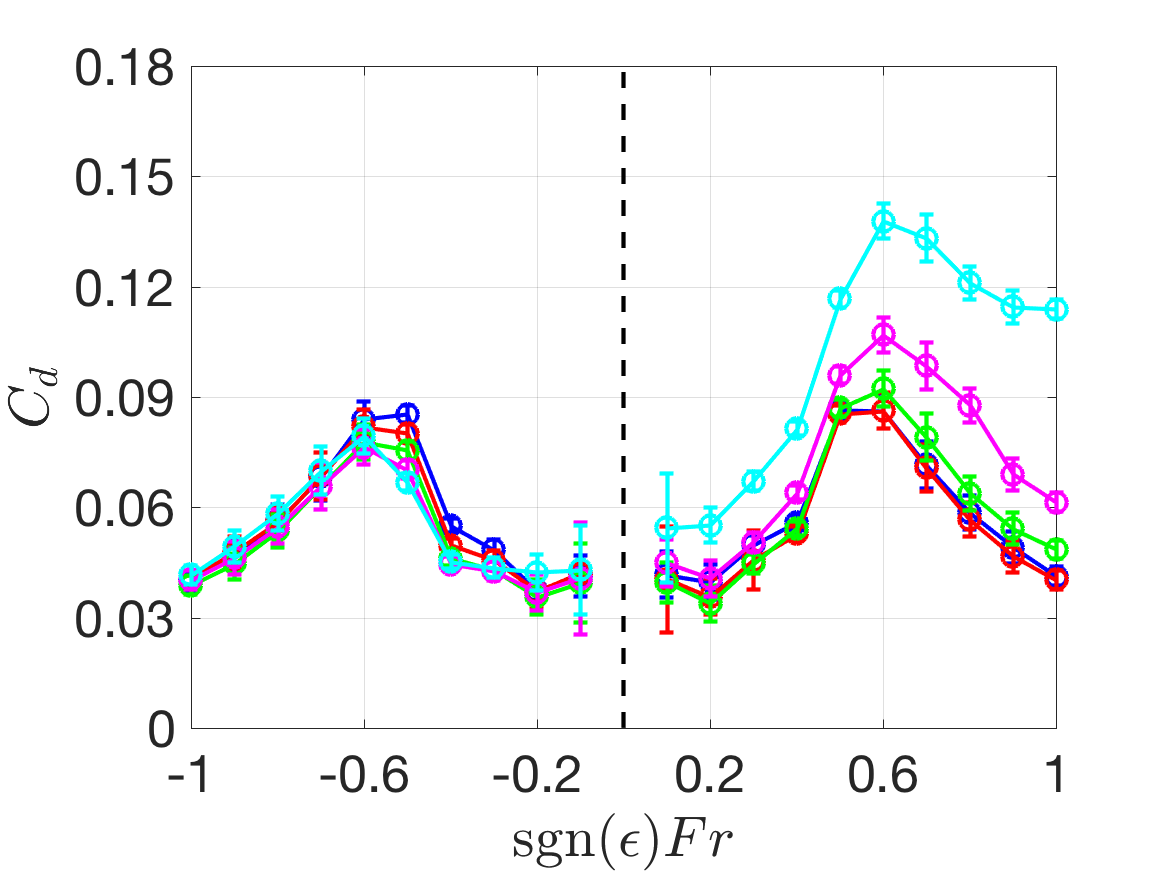}
\put (0,72) {(e)}
\put (42,72) {$\boldsymbol{d=0.5}$}
\end{overpic}
\begin{overpic}[width=0.45\textwidth]{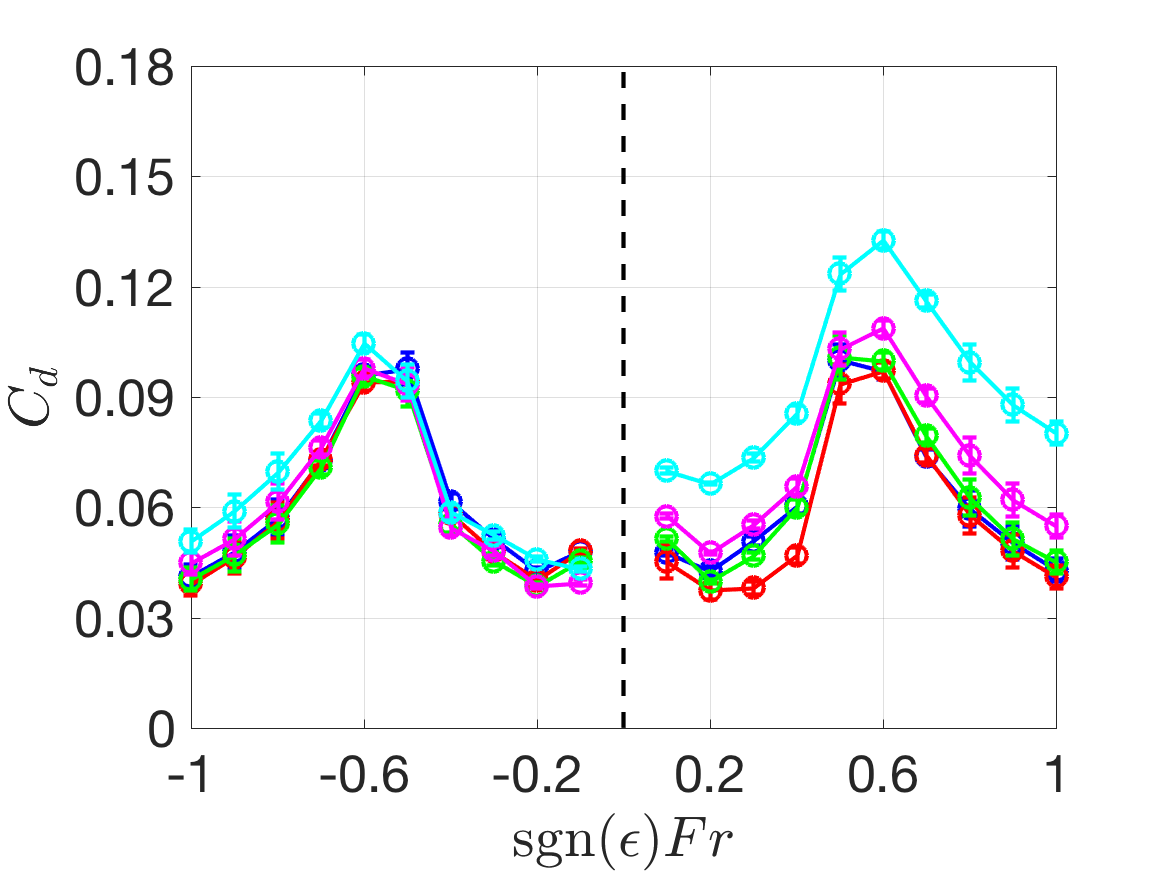}
\put (-30,80) {\large \bf $k$-$\omega$ SST calculations}
\put (0,72) {(f)}
\put (42,72) {$\boldsymbol{d=1.0}$}
\end{overpic}\\
\vspace{0.2cm}
\begin{overpic}[width=0.45\textwidth]{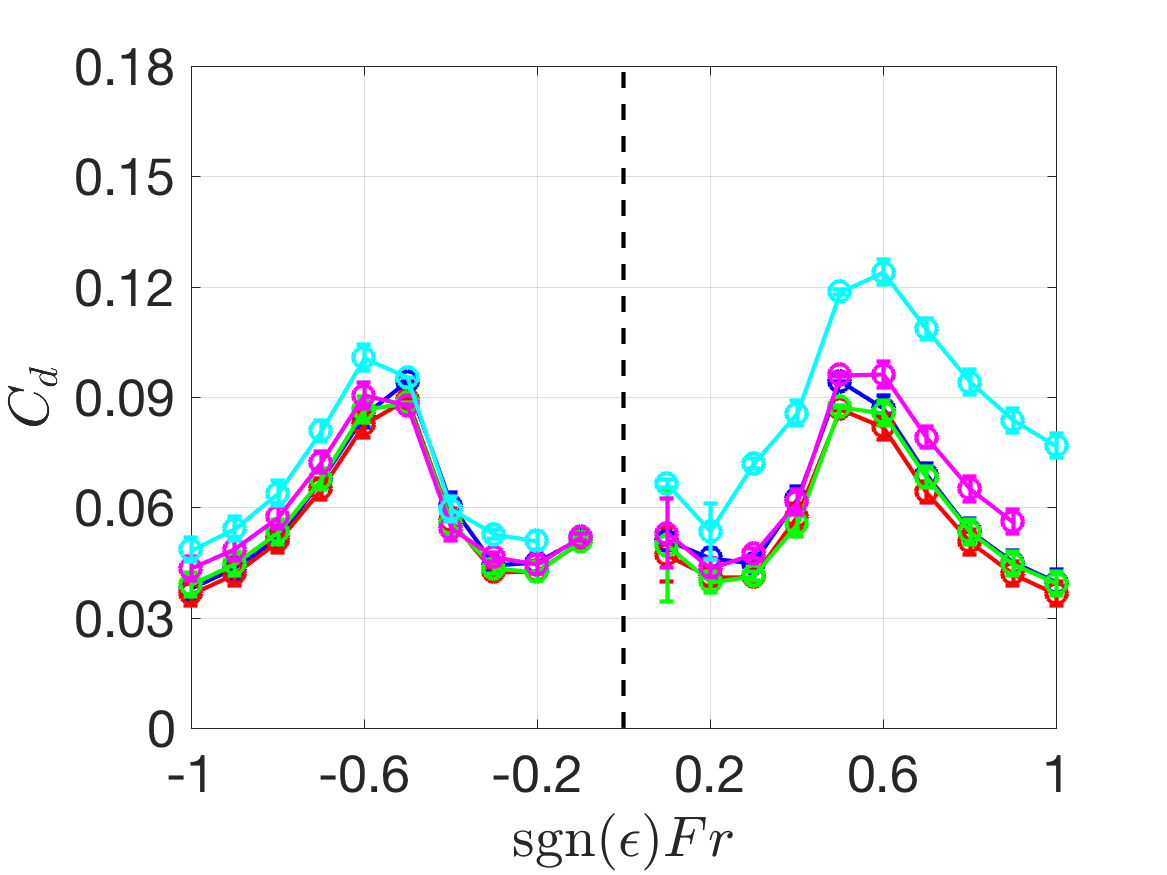}
\put (0,72) {(g)}
\put (42,72) {$\boldsymbol{d=1.25}$}
\end{overpic}
\begin{overpic}[width=0.45\textwidth]{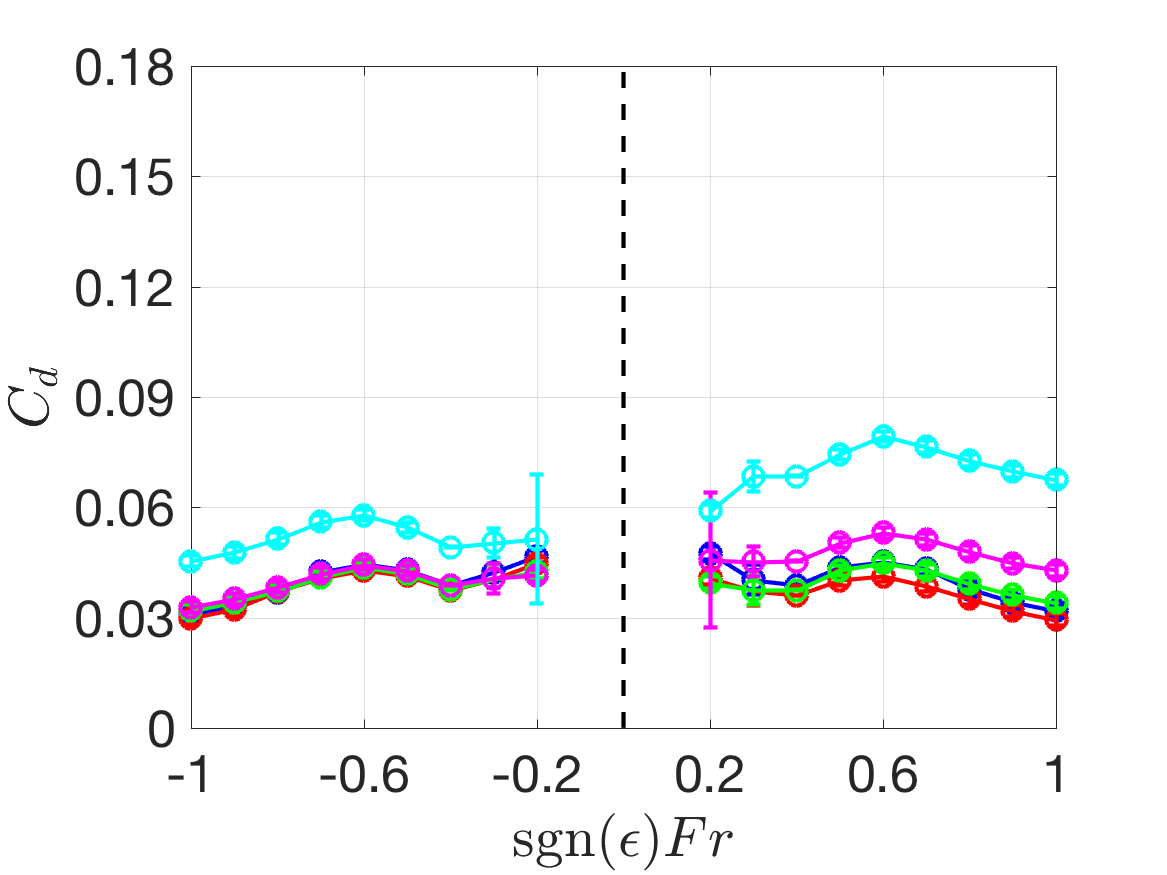}
\put (0,72) {(h)}
\put (42,72) {$\boldsymbol{d=2.0}$}
\end{overpic}
\caption{(a, b, c, d) Additional tow-tank measurements of the drag coefficient $C_d$ for the bluff family of shapes at various different depths. (e, f, g, h) Corresponding calculations of $C_d$ using the $k$-$\omega$ SST model.  \label{app_bluff}}
\end{figure}
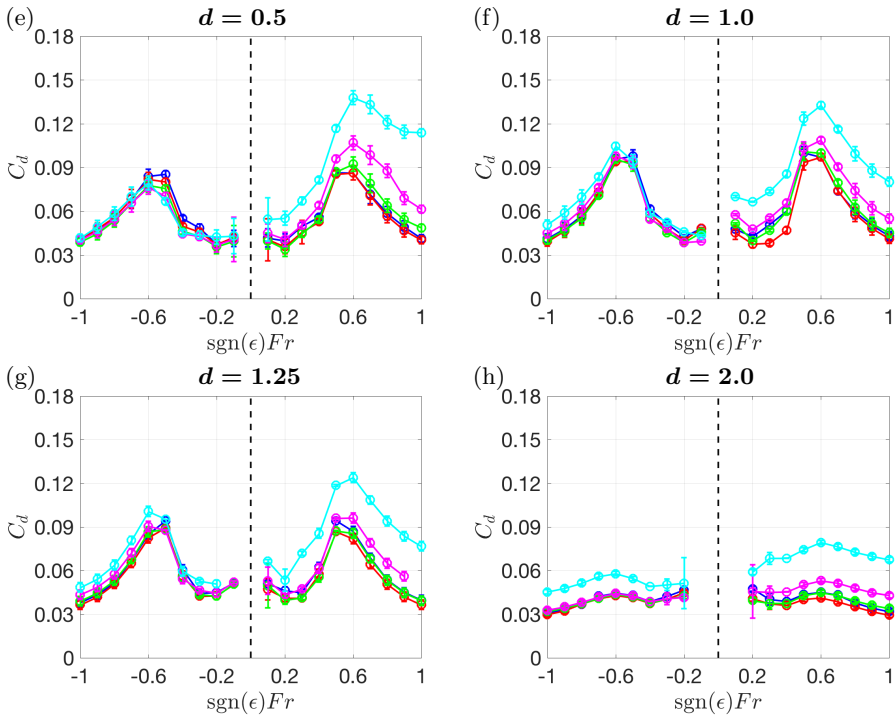

Earlier, in figure \ref{exp}(a), we displayed experimental drag coefficient measurements for depth $d=0.5$ for the hulls from the slender family. Similarly, here in figure \ref{app_slender} we display measurements for depths $d=0.25, 0.75, 1.0$ and for Froude numbers in the range $\mathrm{\it Fr}\in[-1.5,1.5]$. In figure \ref{app_bluff}(a,b,c,d) we also display experimental measurements for the bluff family of shapes at depths $d=0.5, 1.0, 1.25, 2.0$, and for Froude numbers in the range $\mathrm{\it Fr}\in[-1,1]$.
The corresponding $k$-$\omega$ SST calculations of the drag coefficient for the bluff family of shapes are in figure \ref{app_bluff}(e,f,g,h).

\section{Numerical fitting of the turbulent boundary layer}
\label{appC}

In this section we briefly describe the optimisation procedure followed to generate the fitted boundary layer profiles in figure \ref{anal}(a), and the corresponding drag curves in (d).
As explained in Section \ref{sec_exp}, drag measurements have been taken, both using a tow-tank experiment, and in a wind tunnel. In this way, it is possible to isolate the measured wave drag coefficient for a given hull shape $\hat{f}(\hat{x})$, at a given value of the Froude number $\mathrm{\it Fr}$, and the depth of motion $d$. Let us denote the wave drag coefficient derived from this procedure as $C_w^*(\hat{f}(\hat{x}),\mathrm{\it Fr},d)$. 

Now, consider the wave drag coefficient calculated using our modification to Michell's theory (\ref{CwG2}). This theoretical prediction is calculated for a given boundary layer profile $\hat{\delta}_{0.99}(\hat{x})$ and a given hull shape $\hat{f}(\hat{x})$. Hence, we denote the theoretical prediction from (\ref{CwG2}) as $\tilde{C}_w(\hat{\delta}_{0.99}(\hat{x}),\hat{f}(\hat{x}),\mathrm{\it Fr},d)$. In the following numerical fitting procedure, we seek to find the boundary layer profile $\hat{\delta}_{0.99}(\hat{x})$ that, when inserted into (\ref{CwG2}), gives the closest fit possible to the experimentally derived values $C_w^*(\hat{f}(\hat{x}),\mathrm{\it Fr},d)$. 

To perform the fit, we use a least-squares minimisation approach. We run the optimisation for each hull shape $\hat{f}(\hat{x})$ separately. Therefore, for each $\hat{f}(\hat{x})$ we set the objective function as
\beq
J\lb \hat{\delta}_{0.99}(\hat{x})\rb:=\sum_{d\in X_d}\sum_{\mathrm{\it Fr} \in X_{\mathrm{\it Fr}}} \lb \tilde{C}_w(\hat{\delta}_{0.99}(\hat{x}),\hat{f}(\hat{x}),\mathrm{\it Fr},d) - C_w^*(\hat{f}(\hat{x}),\mathrm{\it Fr},d) \rb^2, \label{lsmin}
\eeq
where $X_{\mathrm{\it Fr}}=\{\mathrm{\it Fr}_1,\mathrm{\it Fr}_2,\ldots ,\mathrm{\it Fr}_n\}$ and $X_d=\{d_1,d_2,\ldots ,d_n\}$ denote the set of experimental measurements.

To be physically realistic, we place some constraints on the control function $\hat{\delta}_{0.99}(\hat{x})\in C^\infty [-1/2,1/2]$. Firstly, we require that the boundary layer begins growing at the leading edge of the body, such that
\beq
\hat{\delta}_{0.99}(-1/2)=0.\label{con1}
\eeq
Secondly, we require a non-shrinking boundary layer, such that 
\beq
\hat{\delta}_{0.99}'(\hat{x})\geq 0.\label{con2}
\eeq
Finally, to regularise the optimisation and ensure that the boundary layer profile remains sufficiently smooth, we add a term to the objective function (\ref{lsmin}) that penalises large boundary layer growth rates. Hence, we we replace (\ref{lsmin}) with
\beq
\mathcal{J}\lb\hat{\delta}_{0.99}(\hat{x})\rb:={J}\lb\hat{\delta}_{0.99}(\hat{x})\rb + \mu \int_{-1/2}^{1/2} \hat{\delta}_{0.99}'(\hat{x})^2\,\mathrm{d}\hat{x}.\label{lsmin2}
\eeq
The penalty parameter $\mu$ is chosen to be sufficiently large that regularity is achieved, whilst not being too large that the solution is dramatically affected. For all of the hulls in this study, we have performed a sensitivity analysis on $\mu$ to confirm the stability of the fitted boundary layer profile.

To summarise, the optimisation problem consists of minimising the penalised least squares residual (\ref{lsmin2}), subject to the constraints (\ref{con1}) and (\ref{con2}). 
We solve this numerically, following the same procedure as \citet{benham2018optimal}. This involves discretising the boundary layer shape $\hat{\delta}_{0.99}(\hat{x})$ and treating each of the discretised values as a decision variable. We use the interior point method, with the IpOpt implementation \citep{nocedal2006numerical, wachter2006implementation}. Gradients are calculated using automatic differentiation in the JuMP package \citep{dunning2017jump} of the Julia programming language \citep{bezanson2017julia}.

The resulting fitted boundary layer profiles are plotted in figure \ref{anal}(a) for each hull shape. Then, we insert the boundary layer profiles into (\ref{CwG2}) for Froude numbers in the range $\mathrm{\it Fr}\in[0.3,1.0]$ and at depth $d=0.5$ to calculate the wave drag coefficient $\tilde{C}_w$. These calculations are then added to the wind tunnel measurements $C_f+C_s$ from figure \ref{exp}(b), to produce the total drag coefficient curves $C_d$ in figure \ref{anal}(d).

\bibliographystyle{jfm}
\bibliography{bibfile}

\end{document}